\documentclass[11pt]{article}

\pdfoutput=1

\usepackage[utf8]{inputenc}
\usepackage[top=50pt,bottom=50pt,left=68pt,right=66pt]{geometry}
\usepackage{amsmath}
\usepackage{amssymb}
\usepackage{mathrsfs}
\usepackage{graphicx,caption,subcaption}
\usepackage{multirow}
\usepackage{url}
\newcommand{\overbar}[1]{\mkern1.5mu\overline{\mkern-1.5mu#1\mkern-1.5mu}\mkern 1.5mu}

\newenvironment{rcases}
  {\left.\begin{aligned}}
  {\end{aligned}\right\rbrace}

\def\a{\alpha}

\def\L{\Lambda}

\def\bo{{\raise-.3ex\hbox{\large$\Box$}}}               % D'Alembertian
\def\TH{{\raise.2ex\hbox{$\displaystyle \bigodot$}\mskip-4.7mu \llap H \;}}
\def\face{{\raise.2ex\hbox{$\displaystyle \bigodot$}\mskip-2.2mu \llap {$\ddot
        \smile$}}}                                      % happy face
\def\leftrightarrowfill{$\mathsurround=0pt \mathord\leftarrow \mkern-6mu
        \cleaders\hbox{$\mkern-2mu \mathord- \mkern-2mu$}\hfill
        \mkern-6mu \mathord\rightarrow$}
\def\dvec#1{\vbox{\ialign{##\crcr
        \leftrightarrowfill\crcr\noalign{\kern-1pt\nointerlineskip}
        $\hfil\displaystyle{#1}\hfil$\crcr}}}           % <--> accent
\def\[{\lfloor{\hskip 0.35pt}\!\!\!\lceil}
\def\]{\rfloor{\hskip 0.35pt}\!\!\!\rceil}

\def\low#1{{\raise -3pt\hbox{${\hskip 0.75pt}\!_{#1}$}}}

\newcommand{\be}{\begin{equation}}
\newcommand{\ee}{\end{equation}}
\newcommand{\nbe}{\begin{equation*}}
\newcommand{\nee}{\end{equation*}}

\begin{document}
\thispagestyle{empty}

{\hbox to\hsize{
\vbox{\noindent July 2019 \hfill IPMU19-0096}}}
{\hbox to\hsize{
\vbox{\noindent  }}}
\noindent
\vskip2.0cm

\noindent
\vskip1.0cm
\begin{center}

{\Large\bf Generalized dilaton-axion models of inflation, de Sitter vacua 
\vglue.1in and spontaneous SUSY breaking in supergravity}

\vglue.3in

Yermek Aldabergenov~${}^{a,b,}$\footnote{yermek.a@chula.ac.th}, Auttakit Chatrabhuti~${}^{a,}$\footnote{auttakit.c@chula.ac.th} and Sergei V. Ketov~${}^{c,d,e,}$\footnote{ketov@tmu.ac.jp}
\vglue.1in

${}^a$~Department of Physics, Faculty of Science, Chulalongkorn University,\\
Thanon Phayathai, Pathumwan, Bangkok 10330, Thailand\\
${}^b$~Institute for Experimental and Theoretical Physics, Al-Farabi Kazakh National University, \\
71 Al-Farabi Avenue, Almaty 050040, Kazakhstan \\
${}^c$~Department of Physics, Tokyo Metropolitan University, \\
Minami-ohsawa 1-1, Hachioji-shi, Tokyo 192-0397, Japan \\
${}^d$~Research School for High Energy Physics, Tomsk Polytechnic University,\\
30 Lenin Avenue, Tomsk 634050, Russian Federation \\
${}^e$~Kavli Institute for the Physics and Mathematics of the Universe (IPMU),
\\The University of Tokyo, Chiba 277-8568, Japan \\
\vglue.1in
\end{center}

\vglue.3in

\begin{center}
{\Large\bf Abstract}
\vglue.2in
\end{center}

We propose the unified models of cosmological inflation, spontaneous SUSY breaking, and the dark energy (de Sitter vacuum) in $N=1$ supergravity with the dilaton-axion chiral superfield $T$ in the presence of an $N=1$  vector multiplet with the alternative Fayet-Iliopoulos term. By using the K\"ahler potential as $K=-\alpha\log(T+\overbar{T})$ and the superpotential as a sum of a constant and a linear term, we find that viable inflation is possible for $3\leq\alpha\leq\alpha_{\rm max}$ where $\alpha_{\rm max}\approx 7.235$. Observations of the amplitude of primordial scalar perturbations fix the SUSY breaking scale in our models as high as $10^{13}$ GeV. In the case of $\alpha>3$ the axion gets the tree-level (non-tachyonic) mass comparable to the inflaton mass.

\newpage

\section{Introduction}

Supergravity is well motivated as the possible theoretical interface between (a) high-energy physics
(well) beyond the Standard Model (SM) of elementary particles, (b) gravity beyond the Concordance
($\L$CDM) Cosmological Model, and (c) string theory as the theory of quantum gravity whose low-energy effective action is described by supergravity. A phenomenological  description of high energy particle physics and cosmology in supersymmetry (SUSY) and supergravity is known to be non-trivial, though many viable models exist, see e.g., the reviews \cite{Nilles:1983ge,Aitchison:2007fn,Ketov:2012yz,Ketov:2019mfc} and the references therein. No signs of SUSY at the Large Hadron Collider (LHC) may hint towards a high scale of SUSY phenomena. At such scales (indirect) cosmological probes of SUSY prevail over (direct) experimental probes at particle colliders. The early Universe is, therefore, the natural place for physical applications of supergravity.

A simultaneous description of cosmological inflation and dark energy (as the positive cosmological constant) in supergravity is another challenge due to the huge difference in the relevant scales and the need of (spontaneous) SUSY breaking. The standard approach in supergravity is based on the use of {\it chiral} $N=1$ superfields in four spacetime dimensions with the input given by a K\"ahler potential $K$ and a superpotential $W$. Then the scalar potential and the kinetic terms of the scalar field components are uniquely defined, and the phenomenological model building amounts to choosing both $K$ and $W$ in order to achieve a viable single-field inflation consistent with the Cosmic Microwave Background (CMB) observations and a de Sitter (dS) vacuum after inflation. There are several problems with that approach. First, the input given by $K$ and $W$ allows infinitely many choices. Second, it always leads to the multi-scalar framework so that one has to choose the inflaton direction in the field space and suppress the non-inflaton scalars during inflation in order to prevent spoiling of the inflaton slow roll and get enough number of e-foldings. Third, after inflation one has to get the hierarchy between the (high) SUSY breaking scale allowing large masses for the superpartners of the SM particles and the (low) dark energy scale given by the cosmological constant. Getting that hierarchy may require {\it two} different mechanisms of spontaneous SUSY breaking. 

It is possible to reduce (and minimize) the number of scalars in the inflationary models by employing a massive (irreducible) $N=1$ {\it vector} multiplet as the inflaton supermultiplet, instead of a chiral one~\cite{Farakos:2013cqa,Ferrara:2013kca}.  The massive vector multiplet has only {\it one} (real) physical scalar that can be identified with inflaton, while its fermionic superpartner can be identified with goldstino in the minimalistic setup for inflation in supergraviity (cf. Refs.~\cite{Ketov:2014hya,Ketov:2014qha}). To avoid SUSY restoration after inflation in a Minkowski vacuum (it was the drawback of the first supergravity models with inflaton in a vector multiplet), one may either add the hidden sector described by a chiral (Polonyi) superfield as in Refs.~\cite{Aldabergenov:2016dcu,Aldabergenov:2017bjt,Addazi:2017ulg} or introduce the alternative (new)  Fayet-Iliopoulos (FI) terms as in Refs.~\cite{Aldabergenov:2017hvp,Aldabergenov:2018nzd}.~\footnote{The alternative FI terms without gauging the R-symmetry were introduced in Refs.~\cite{Cribiori:2017laj,Kuzenko:2018jlz}.} Moreover, one can also combine both approaches and derive the supergravity-based inflationary models with inflaton in a massive vector multiplet in the presence of the FI term, with both F-type and D-type SUSY breaking needed for the hierarchy of scales \cite{Abe:2018plc,Abe:2018rnu}. In all those cases, the canonical K\"ahler potential and a linear superpotential for Polonyi superfield were chosen, like the original Polonyi model \cite{Polonyi:1977pj}.

Another approach is based on the use of the "dilaton-axion" superfield $T$ by replacing the canonical (free) K\"ahler potential by the generalized "no-scale" one as follows \cite{Ellis:1983ei}:
\begin{equation}
   K=-\alpha\log(T+\overbar{T})~,\label{Kahler_intro}
\end{equation}
The corresponding $N=1$ non-linear sigma-model has the $SL(2;\mathbb{R})/SO(2)$ (or $SU(1,1;\mathbb{C})/U(1)$) tangent space of K\"ahler curvature $R_K=2/\alpha$, and is of particular interest for particle phenomenology because  such K\"ahler potential in the case of 
$\a=3$  arises in generic heterotic string compactifications and allows for the realistic particle model building \cite{Ellis:1983ei,Cremmer:1983bf,Ellis:1983sf,Ellis:1984bm}. Then $T$ can be identified with the volume modulus of the compactified manifold in heterotic string theory. It is remarkable that the same K\"ahler potential with $\alpha=3$ also arises in the modified $F(R)$ supergravity after its dualization \cite{Cecotti:1987sa,Gates:2009hu,Ketov:2009sq}. 

It is, however, also known that the case of $\a=3$ in Eq.~(\ref{Kahler_intro}) with just a {\it single}
chiral superfield is not viable for cosmological applications because it does not allow stable dS vacua and cannot be used for realizing Starobinsky inflation \cite{Starobinsky:1980te} with {\it any} choice of the superpotential $W$ \cite{Ketov:2012yz,Ellis:2015kqa,Ellis:2015xna,Ellis:2017xwz,Ellis:2018xdr}, although there are single field models with generalized $\alpha$ ($\alpha$-attractors) leading to a supersymmetric Minkowski vacuum \cite{Kallosh:2013yoa,Roest:2015qya,Linde:2015uga}.

The "no-scale" supergravity  was  successfully used for describing inflation in Refs.~\cite{Ellis:2015kqa,Ellis:2015xna,Ellis:2017xwz,Ellis:2018xdr,Ellis:2018zya,Ellis:2019bmm} with the help of at least {\it two chiral} superfields and the K\"ahler potential 
\begin{equation}
   K=-3\log\left(T+\overbar{T}-\sum_{i=1}^{p-1}  |\Phi^i|^2/3\right)~,\label{Kahler_two}
\end{equation}
where $T$ is the volume modulus, and $\Phi^i$ are the matter chiral superfields parametrizing the
non-linear sigma-model tangent space  $SU(p,1;\mathbb{C})/SU(p)\times U(1)$, with the suitable superpotential.

In this paper we use a single  "dilaton-axion" chiral superfield $T$ with the K\"ahler potential  (\ref{Kahler_intro}) but introduce a single {\it vector} multiplet in addition. We demonstrate that it leads
to the viable set of cosmological models describing inflation, dS vacua and spontaneous SUSY breaking.   

We recall that the original Starobinsky model of inflation \cite{Starobinsky:1980te} is based on the modified $(R+R^2)$ gravity, while its extension in the new-minimal formulation of supergravity has  the {\it dual} description in terms of the standard supergravity coupled to a massive vector multiplet or, equivalently, a massless vector multiplet and a St\"uckelberg chiral multiplet with the K\"ahler potential \cite{Ferrara:2013kca}
\begin{equation}
    K=-3\log(T+\overbar{T})+3(T+\overbar{T})~.
\end{equation}
The last term can be identified with the FI term of the gauged R-symmetry 
(in a non-R-symmetric frame), because the D-term of this model results in the Starobinsky potential
 \cite{Farakos:2013cqa,Ferrara:2013kca,Ferrara:2013rsa}.
 
  The authors of Ref.~\cite{Antoniadis:2016aal} studied even more general models of a single chiral multiplet and an abelian vector multiplet  with the gauged R-symmetry and the K\"ahler potential having two parameters $\alpha$ and $\beta$,  
\begin{equation}
    K=-\alpha\log(T+\overbar{T})+\beta(T+\overbar{T})~,
\end{equation}
and found that slow-roll inflation consistent with observations is only possible for $\alpha=1,2$ after adding some non-perturbative corrections. SUSY is spontaneously broken after inflation in those models, with the gravitino mass in the TeV range.

In this paper we find that a non-vanishing D-term allows us to introduce the new inflationary models based on the K\"ahler potentials having the form \eqref{Kahler_intro} with a single chiral superfield and 
a single vector superfield. The other examples of the D-term
based on the alternative FI terms \cite{Cribiori:2017laj,Kuzenko:2018jlz} can be found in 
Refs.~\cite{Aldabergenov:2018nzd,Farakos:2018sgq,Antoniadis:2018oeh,Cribiori:2018dlc,Antoniadis:2019hbu}. Those FI terms provide a tunable positive cosmological constant 
or dS uplifting of the vacuum after inflation \cite{Aldabergenov:2017hvp,Abe:2018plc,Abe:2018rnu,Antoniadis:2018cpq}. Our inflationary models in this paper have the K\"ahler potential \eqref{Kahler_intro} and the Polonyi-type linear superpotential (without gauging the shift symmetry of $K$) leading to the spontaneous  F-type SUSY breaking. In addition, the simplest alternative FI term leads to another D-type SUSY breaking and uplifts an Anti-dS (AdS) minimum of the F-term scalar potential to a dS minimum.

The paper is organized as follows. Our setup is given in Sec.~2. In Sec.~3 we study vacua and
SUSY breaking. In Sec.~4 we study inflation in our framework and analyze in detail the models with integer $\alpha$. In particular, we derive the explicit values of the dilaton and axion masses and the SUSY breaking parameters by fixing the inflationary observables with the CMB observational data. We  conclude in Sec.~5. The basic formulae about the standard $N=1$ supergravity and the alternative FI term are given in Appendix. We set the reduced Planck mass as $M_{\rm Pl}=\kappa^{-1}=1$ unless otherwise stated.

\section{The setup}

Let us consider the following K\"ahler potential and the superpotential:
\begin{gather}
    K=-\alpha\log(T+\overbar{T})~,\label{Kahler0}\\
    W=\lambda+\mu T~,\label{superp0}
\end{gather}
where $\alpha$ is a positive real constant, $\lambda$ and $\mu$ are complex parameters. The 
$T$ is parametrized as
\begin{equation}
    T=e^{-\sqrt{\frac{2}{\alpha}}\phi}+it \label{parametrize}
\end{equation}
in terms of the canonical inflaton $\phi$ and its axionic partner $t$. The F-term scalar potential reads
\begin{align}
    V_F &= e^K\left[K^{T\overbar{T}}(W_T+K_TW)(\overbar{W}_{\overbar{T}}+K_{\overbar{T}}\overbar{W})-3|W|^2\right]=\nonumber\\
    &=\frac{\alpha-3}{2^\alpha}(|\lambda|^2+\omega_2t+|\mu|^2t^2)e^{\sqrt{2\alpha}\phi}+\frac{(\alpha-5)\omega_1}{2^\alpha}e^{(\alpha-1)\sqrt{\frac{2}{\alpha}}\phi}+\frac{(\alpha^2-7\alpha+4)|\mu|^2}{2^\alpha\alpha}e^{(\alpha-2)\sqrt{\frac{2}{\alpha}}\phi}~,\label{VF0}
\end{align}
where we have used the parametrization \eqref{parametrize} and the notation
\begin{align}
    \omega_1&\equiv \overbar{\lambda}\mu+\lambda\overbar{\mu}=2\lambda^{}_R\mu^{}_R+2\lambda^{}_I\mu^{}_I~,\label{omega1}\\
    \omega_2&\equiv i(\overbar{\lambda}\mu-\lambda\overbar{\mu})=2\lambda^{}_I\mu^{}_R-2\lambda^{}_R\mu^{}_I~.\label{omega2}
\end{align}
The subscripts $R,I$ stand for the real and imaginary parts. It is convenient to trade the complex parameter $\lambda$ for the two real ones, $\omega_1$ and $\omega_2$ defined above.

A generic vacuum of the F-term potential \eqref{VF0} is AdS. However, after introducing an abelian vector multiplet with the simplest alternative FI term \cite{Cribiori:2017laj,Kuzenko:2018jlz} and 
eliminating the auxiliary field $(D)$ of the vector multiplet, one gets a positive contribution
\begin{equation}
    V_D=\frac{g^2\xi^2}{2} \label{VD0}
\end{equation}
to the cosmological constant, where $g$ is the gauge coupling, and $\xi$ is the real FI constant.~\footnote{The model defined by Eqs.~\eqref{Kahler0} and \eqref{superp0} in the presence of the alternative FI term and the linear gauge kinetic function $f(g)=T$ leads to the {\it vanishing} scalar potential (see Eqs.~\eqref{VF0} and \eqref{VD0}) when $\alpha=3$ and $\omega_1=0$ \cite{Aldabergenov:2019hvl}, which is the defining property of the "no-scale" supergravity.}  More details about the alternative FI term and the bosonic action of the standard $N=1$ supergravity can be found in Appendix.

\section{Vacua and SUSY breaking}

Let us analyze minima of the scalar potential (\ref{VF0}). The vacuum equations for the critical points 
and the critical value $V_0$ of the potential read
\begin{align}
    V_0&=Ax^\alpha+Bx^{\alpha-1}+Cx^{\alpha-2}+\frac{g^2\xi^2}{2}~,\label{V0general}\\
    V_x&=\alpha Ax^{\alpha-1}+(\alpha-1)Bx^{\alpha-2}+(\alpha-2)Cx^{\alpha-3}=0~,\label{Vxgeneral}\\
    V_t&=\frac{\alpha-3}{2^\alpha}(\omega_2+2|\mu|^2t_0)x^\alpha=0~,\label{Vtgeneral}
\end{align}
where we have used the notation
\begin{gather}
    x\equiv e^{\sqrt{\frac{2}{\alpha}}\phi_0}~~~{\rm with}~~~\phi_0\equiv\langle\phi\rangle~,\nonumber\\
    A\equiv\frac{\alpha-3}{2^\alpha}(|\lambda|^2+\omega_2t_0+|\mu|^2t_0^2)~,\label{defABCx}\\
    B\equiv\frac{(\alpha-5)\omega_1}{2^\alpha}~,~~~C\equiv\frac{(\alpha^2-7\alpha+4)|\mu|^2}{2^\alpha\alpha}\nonumber~,
\end{gather}
and $V_x\equiv\partial V/\partial x$.

The special value $\alpha=3$ yields the identically vanishing $A$ and $V_t$, thus making the potential $t$-independent. In the next Section we consider separately the case of $\alpha=3$ and then turn to $\alpha\neq 3$. When $\alpha\neq 3$, we can use the solution to Eq. \eqref{Vtgeneral}
as $t_0=-\omega_2/(2|\mu|^2)$ and rewrite $A$ as
\begin{equation}
    A=\frac{\alpha-3}{2^\alpha}\left(|\lambda|^2-\frac{\omega_2^2}{4|\mu|^2}\right)=\frac{(\alpha-3)\omega_1^2}{2^{\alpha+2}|\mu|^2}~,\label{Amin}
\end{equation}
where in the last equation we have used the definitions \eqref{omega1} and \eqref{omega2}. Since $\omega_1$ is real, $A$ becomes negative when $\alpha<3$. Given negative $A$, the potential \eqref{V0general} is unbounded from below because $A$ multiplies the highest power of $x$ 
(for $\alpha<3$ the potential also becomes unstable in the $t$-direction). Therefore, we restrict ourselves to $\alpha\geq 3$ in what follows.

\subsection{The case $\alpha=3$}

Given $\alpha=3$, the scalar potential takes the simple form
\begin{equation}
    V=-\frac{\omega_1}{4}e^{\sqrt{\frac{8}{3}}\phi}-\frac{|\mu|^2}{3}e^{\sqrt{\frac{2}{3}}\phi}+\frac{g^2\xi^2}{2}~,\label{V_alpha=3}
\end{equation}
and has a minimum at
\begin{align}
    \phi_0&=\sqrt{\frac{3}{2}}\log\left(-\frac{2|\mu|^2}{3\omega_1}\right)~,\label{phi0_alpha=3}\\
    V_0&=\frac{g^2\xi^2}{2}+\frac{|\mu|^4}{9\omega_1}~.\label{V0_alpha=3}
\end{align}
The minimum exists only if $\omega_1<0$. The minimum is AdS, Minkowski, or dS, depending on the following relations:
\begin{align}
    g^2\xi^2<\frac{2|\mu|^4}{9|\omega_1|}~&\longrightarrow~{\rm AdS}~,\label{adscond}\\
    g^2\xi^2=\frac{2|\mu|^4}{9|\omega_1|}~&\longrightarrow~{\rm Minkowski}~,\label{minkcond}\\
    g^2\xi^2>\frac{2|\mu|^4}{9|\omega_1|}~&\longrightarrow~{\rm dS}~.\label{dscond}
\end{align}
Hence, by fine-tuning the parameters we can obtain a small positive cosmological constant $V_0$ for the realistic phenomenology.

Defining $\varphi\equiv\phi-\phi_0$ as excitation of the inflaton around its Vacuum Expectation Value (VEV), the potential \eqref{V_alpha=3} can be brought to the form (after using Eq.~\eqref{V0_alpha=3} to eliminate $g\xi$ in terms of $V_0$)
\begin{equation}
    V=V_0+\frac{|\mu|^4}{9|\omega_1|}\left(e^{\sqrt{\frac{2}{3}}\varphi}-1\right)^2
\end{equation}
that gives the realization of the Starobinsky inflationary model (with the cosmological constant) in our framework. The potential is $t$-flat.

SUSY is spontaneously broken by the constant non-vanishing D-term, $D=\langle D \rangle=g\xi$, while $\langle F_T\rangle$ and the gravitino mass are given by
\begin{align}
    \langle F_T\rangle&=\langle-e^{K/2}K^{T\overbar{T}}(\overbar{W}_{\overbar{T}}+K_{\overbar{T}}\overbar{W})\rangle=-\frac{i~{\rm sgn}(\mu)}{\sqrt{12|\omega_1|}}(\omega_2+2|\mu|^2t_0)~,\\
     m_{3/2}^2&=\langle e^K|W|^2\rangle=|-2\omega_1+i(\omega_2+2|\mu|^2t_0)|^2\frac{|\mu|^4}{108|\omega_1|^3}~.
\end{align}
Though $\langle F_T\rangle$ is arbitrary (and may even vanish), the gravitino mass is bounded from below,
\begin{equation}
    m_{3/2}^2\geq\frac{|\mu|^4}{27|\omega_1|}~.\label{alpha=3mm}
\end{equation}

The mass of the inflaton is
\begin{equation}
    m^2_{\varphi}=\frac{4|\mu|^4}{27|\omega_1|}~,
\end{equation}
so that we have the relation $2m_{3/2}\geq m_{\varphi}$.

The bosonic sector also includes a massless axion $t$ and a massless vector. The vector can be
made massive via (additional) super-Higgs effect. A massless scalar is phenomenologically problematic, but the mass of $t$ may be generated either by quantum corrections when $\alpha=3$ (as is usually assumed in the "no-scale" supergravity models), or already at the tree level when $\alpha>3$, as we are going to show in the next Subsection.

\subsection{The case $\alpha>3$: vacuum solutions}

If the axion field is fixed at its VEV, $t_0=-\omega_2/(2|\mu|^2)$, we can rewrite the scalar potential \eqref{VF0} and \eqref{VD0} for $\alpha>3$ as
\begin{equation}
     V=\frac{(\alpha-3)\omega_1^2}{2^{\alpha+2}|\mu|^2}e^{\sqrt{2\alpha}\phi}+\frac{(\alpha-5)\omega_1}{2^\alpha}e^{(\alpha-1)\sqrt{\frac{2}{\alpha}}\phi}+\frac{(\alpha^2-7\alpha+4)|\mu|^2}{2^\alpha\alpha}e^{(\alpha-2)\sqrt{\frac{2}{\alpha}}\phi}+\frac{1}{2}g^2\xi^2~,\label{V_alpha>3}
\end{equation}
assuming $\omega_1\neq 0$. The vacuum equations \eqref{V0general} and \eqref{Vxgeneral} then take the form
\begin{align}
    V_0&=\frac{(\alpha-3)\omega_1^2}{2^{\alpha+2}|\mu|^2}x^\alpha+\frac{(\alpha-5)\omega_1}{2^\alpha}x^{\alpha-1}+\frac{(\alpha^2-7\alpha+4)|\mu|^2}{2^\alpha\alpha}x^{\alpha-2}+\frac{1}{2}g^2\xi^2~,\label{V0_alpha>3}\\
    V_x&=\frac{\alpha(\alpha-3)\omega_1^2}{2^{\alpha+2}|\mu|^2}x^{\alpha-1}+\frac{(\alpha-1)(\alpha-5)\omega_1}{2^\alpha}x^{\alpha-2}+\frac{(\alpha-2)(\alpha^2-7\alpha+4)|\mu|^2}{2^\alpha \alpha}x^{\alpha-3}=0~,\label{Vx_alpha>3}
\end{align}
where $x\equiv e^{\phi_0/\sqrt{2}}$ as before. Eq.~\eqref{Vx_alpha>3} has two solutions,
\begin{equation}
    x_+=\frac{2(-\alpha^2+7\alpha-4)|\mu|^2}{\alpha(\alpha-3)\omega_1}~,~~~x_-=\frac{2(2-\alpha)|\mu|^2}{\alpha\omega_1}~,\label{xpmsolutions}
\end{equation}
that we parametrize as
\begin{equation}
    x_{\pm}=\gamma_{\pm}\frac{|\mu|^2}{\omega_1}~~\begin{cases}
    \gamma_+\equiv\frac{2(-\alpha^2+7\alpha-4)}{\alpha(\alpha-3)}~,\\
    \gamma_-\equiv\frac{2(2-\alpha)}{\alpha}~.
    \end{cases}\label{gammapm}
\end{equation}

\begin{figure}[t]
\centering
\includegraphics[scale=0.7]{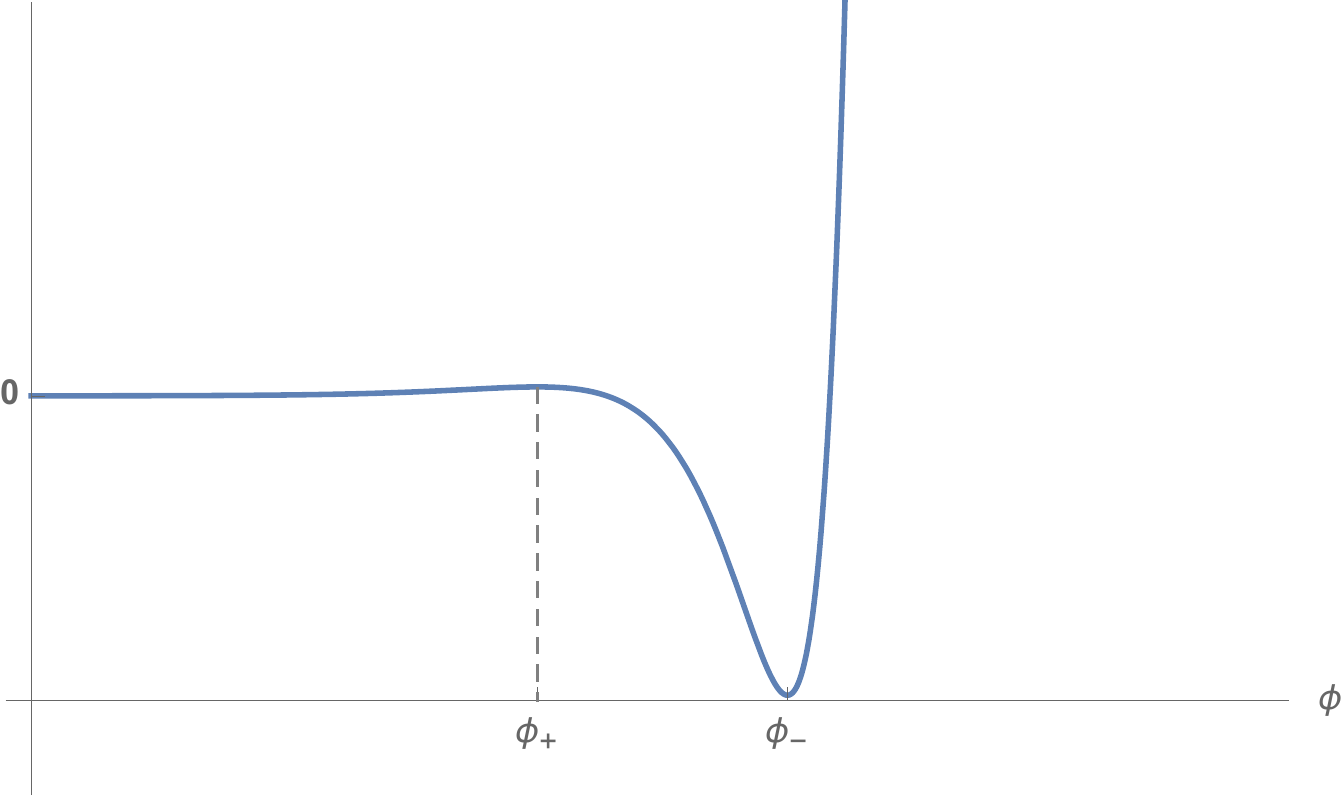}
\captionsetup{width=.8\linewidth}
\caption{The form of the scalar potential \eqref{V_alpha>3} for $\alpha>\frac{1}{2}(7+\sqrt{33})$, $\xi=0$, and negative $\omega_1$. The $\phi_{\pm}$ are defined as $x_{\pm}= e^{\phi_{\pm}/\sqrt{2}}$, where $\phi_-=\phi_0$ is the VEV of $\phi$.}\label{FigVminmax}
\end{figure}

The positivity of $x$ requires $\gamma/\omega_1$ to be positive. Since we have $\alpha>3$, 
the $\gamma_-$ is always negative, while the sign of $\gamma_+$ depends on the choice of $\alpha$. More specifically, we find~\footnote{$\alpha=\frac{1}{2}(7+\sqrt{33})$ is one of the two roots of the polynomial $\alpha^2-7\alpha+4$ that yields $\gamma_+=x_+=0$. Another root is $\alpha=\frac{1}{2}(7-\sqrt{33})<3$ so that  it is excluded from the analysis.}
\begin{equation}
    3<\alpha<\frac{1}{2}(7+\sqrt{33})~\longrightarrow~\gamma_+>0~.
\end{equation}
Hence, if $\omega_1>0$, the $x_+$ should be used as the vacuum solution in this parameter region. On the other hand,
\begin{equation}
    \alpha=\frac{1}{2}(7+\sqrt{33})~\longrightarrow~\gamma_+=0~,
\end{equation}
that invalidates the $x_+$ as a stable solution for the given value of $\alpha$. Thus $\omega_1$ must be negative and $x_-$ should be used as the minimum. Moreover, 
\begin{equation}
    \alpha>\frac{1}{2}(7+\sqrt{33})~\longrightarrow~\gamma_+<0~.\label{alpha>6.4}
\end{equation}
This means that $\omega_1$ should be negative. In this case, both $x_+$ and $x_-$ (i.e. $\phi_+$ and $\phi_-$) are the valid stationary points and, in fact, the extrema -- not inflection points -- because
\begin{equation}
    \left.\frac{\partial^2 V(\phi)}{\partial \phi^2}\right|_{\phi=\phi_{\pm}}\neq 0~,
\end{equation}
where $V(\phi)$ is given by Eq.~\eqref{V_alpha>3}. Setting $\xi=0$ and using $\omega_1=-|\omega_1|$, we get 
\begin{align}
    V_0|_{\phi=\phi_+}&=\frac{|\mu|^{2(\alpha-1)}}{2^{\alpha+2}\alpha|\omega_1|^{\alpha-2}}\left(\frac{\alpha^2-7\alpha+4}{\alpha(\alpha-3)}\right)^{\alpha-1}>0~,\\
    V_0|_{\phi=\phi_-}&=-\frac{3|\mu|^{2(\alpha-1)}}{2^{\alpha+2}|\omega_1|^{\alpha-2}}\left(\frac{1-\frac{2}{\alpha}}{(\alpha-2)^2}\right)<0~,
\end{align}
for $\alpha>(7+\sqrt{33})/2$. It means that $\phi_+$ is a local maximum, while $\phi_-$ is a global minimum. The general form of the potential is shown in Fig.~\ref{FigVminmax}. The existence of the local maximum at $\phi_+$ means that the potential is of the hilltop-type and, therefore, we should consider inflation in the cases $3\leq\alpha\leq(7+\sqrt{33})/2$ and $\alpha>(7+\sqrt{33})/2$ separately. Since the value of $\alpha=(7+\sqrt{33})/2$ is special, we  introduce the notation
\begin{equation}
    \alpha_*\equiv\frac{1}{2}(7+\sqrt{33})\approx 6.37~.
\end{equation}

It is noteworthy that the choice of $\alpha=5$ leads to $\gamma_+=-\gamma_-=6/5$, so that the scalar potentials in the cases $\omega_1>0$ and $\omega_1<0$ exactly coincide.

\subsection{The case $\alpha>3$: SUSY breaking and scalar masses}

In the case of $3<\alpha\leq\alpha_*$ we use again the axion VEV, $t_0=-\omega_2/(2|\mu|^2)$, and the general solution $x=\gamma|\mu|^2/\omega_1$ to find $F_T$ and the gravitino mass at the minimum,
\begin{align}
    \langle F_T\rangle&=\frac{(\gamma\alpha+2\alpha-4)|\mu|^{2-\alpha}}{2^{\alpha/2}\alpha\mu}\left(\frac{\gamma}{\omega_1}\right)^{\frac{\alpha}{2}-2}~,\\
    m^2_{3/2}&=\frac{(\gamma+2)^2|\mu|^{2(\alpha-1)}}{4\cdot 2^\alpha}\left(\frac{\gamma}{\omega_1}\right)^{\alpha-2}~.
\end{align}
Substituting the two solutions  from Eq.~\eqref{gammapm} yields
\begin{align}
    \langle F_T\rangle|_{x_+}&=\frac{(\alpha+1)|\mu|^{\alpha-2}}{\alpha^{\frac{\alpha}{2}-1}(\alpha-3)\mu}\left[\frac{\alpha^2-7\alpha+4}{(3-\alpha)\omega_1}\right]^{\frac{\alpha}{2}-2}~,\label{FTx+}\\
    \langle F_T\rangle|_{x_-}&=0~,\label{FTx-}
\end{align}
while the gravitino mass in both cases ($x_+$ and $x_-$) is non-vanishing. Now recall that for $3<\alpha<\alpha_*$ the vacuum solution is $x_+$ if $\omega_1$ is positive, and $x_-$ if $\omega_1$ is negative, while for $\alpha\geq\alpha_*$ only $x_-$ can be a stable vacuum solution and this requires a negative $\omega_1$. Thus, we conclude that if $3<\alpha<\alpha_*$, a positive $\omega_1$ leads to the mixed F- and D-term SUSY breaking, while a negative $\omega_1$ leads to the pure D-term SUSY breaking. If $\alpha\geq\alpha_*$, only the pure D-term breaking is possible (we exclude runaway solutions).

As for the mass of the axion $t$, we first get
\begin{equation}
    m_t^2=\frac{(\alpha-3)|\mu|^2}{2^{\alpha-1}}e^{\sqrt{2\alpha}\phi_0}~.
\end{equation}
However, the $t$ is not canonical at the $\phi$-minimum because the $\phi_0$ is non-vanishing and
\begin{equation}
    e^{-1}{\cal L}_{\rm kin}(t,\phi_0)=-\frac{\alpha}{4}e^{\sqrt{\frac{8}{\alpha}}\phi_0}(\partial t)^2~,
\end{equation}
as can be seen from Eq.~\eqref{compL} in Appendix. Though it is impossible to canonically normalize
the kinetic term of $t$ for all values of $\phi$, it is certainly possible at the reference point $\phi_0$ 
by the rescaling 
\begin{equation}
t=\sqrt{\frac{2}{\alpha}}e^{-\sqrt{\frac{2}{\alpha}}\phi_0}t'~,
\end{equation}
where $t'$ is the "canonical" axion. Its mass squared is then given by
\begin{equation}
    m_{t'}^2=\frac{(\alpha-3)|\mu|^2}{2^{\alpha-2}\alpha}e^{(\alpha-2)\sqrt{\frac{2}{\alpha}}\phi_0}=\frac{(\alpha-3)|\mu|^{2(\alpha-1)}}{2^{\alpha-2}\alpha}\left(\frac{\gamma}{\omega_1}\right)^{\alpha-2}~,\label{tprimemass}
\end{equation}
where we have used the general vacuum solution $e^{\sqrt{\frac{2}{\alpha}}\phi_0}\equiv x=\gamma|\mu|^2/\omega_1$. 

The inflaton mass can be read off from Eq.~\eqref{V_alpha>3} after using $\varphi=\phi-\phi_0$ and substituting the general $x$ solution \eqref{gammapm}. We find
\begin{equation}
    m_\varphi^2=2\left(\frac{\gamma}{2}\right)^\alpha\left[\frac{\alpha(\alpha-3)}{4}+\frac{(\alpha-5)(\alpha-1)^2}{\alpha\gamma}+\frac{(\alpha^2-7\alpha+4)(\alpha-2)^2}{\alpha^2\gamma^2}\right]\frac{|\mu|^{2(\alpha-1)}}{\omega_1^{\alpha-2}}~.\label{mvarphi^2}
\end{equation}
It is convenient to define the mass ratios
\begin{equation}
    \Delta_\pm\equiv\left.\frac{m_{t'}}{m_\varphi}\right|_{\gamma=\gamma_\pm}~,~~~\Gamma_\pm\equiv\left.\frac{m_{3/2}}{m_\varphi}\right|_{\gamma=\gamma_\pm}~,
\end{equation}
where the $\gamma_{\pm}$ are defined in Eq.~\eqref{gammapm}. The parameters $\mu$ and $\omega_1$ cancel out in $\Delta$ and $\Gamma$ that can be readily plotted as the functions of $\alpha$.

\begin{figure}
\centering
\begin{subfigure}{.5\textwidth}
  \centering
  \includegraphics[width=.98\linewidth]{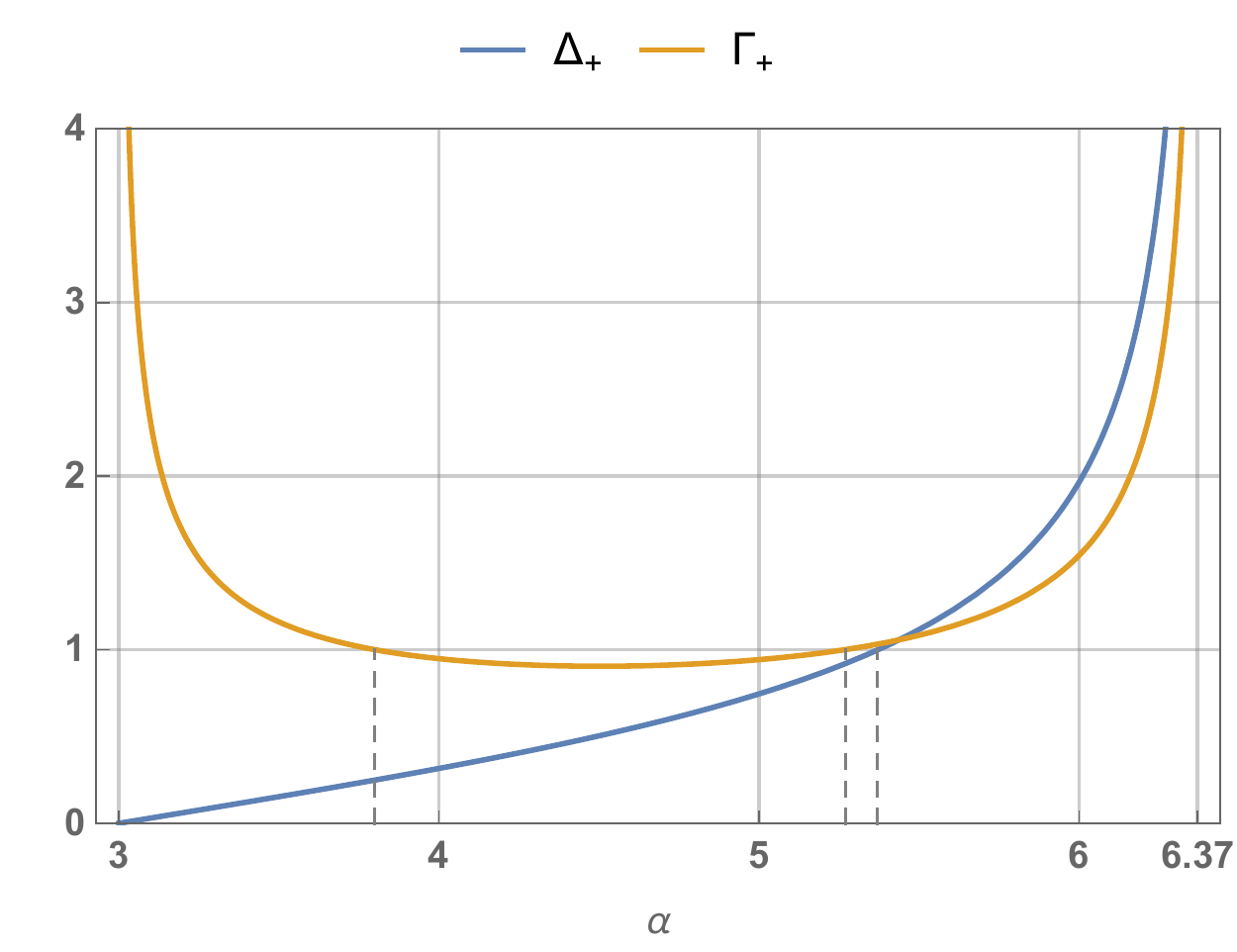}
  \caption{$\Delta_+$ (blue curve) and $\Gamma_+$ (orange curve).}
  \label{FigDGp}
\end{subfigure}%
\begin{subfigure}{.5\textwidth}
  \centering
  \includegraphics[width=1\linewidth]{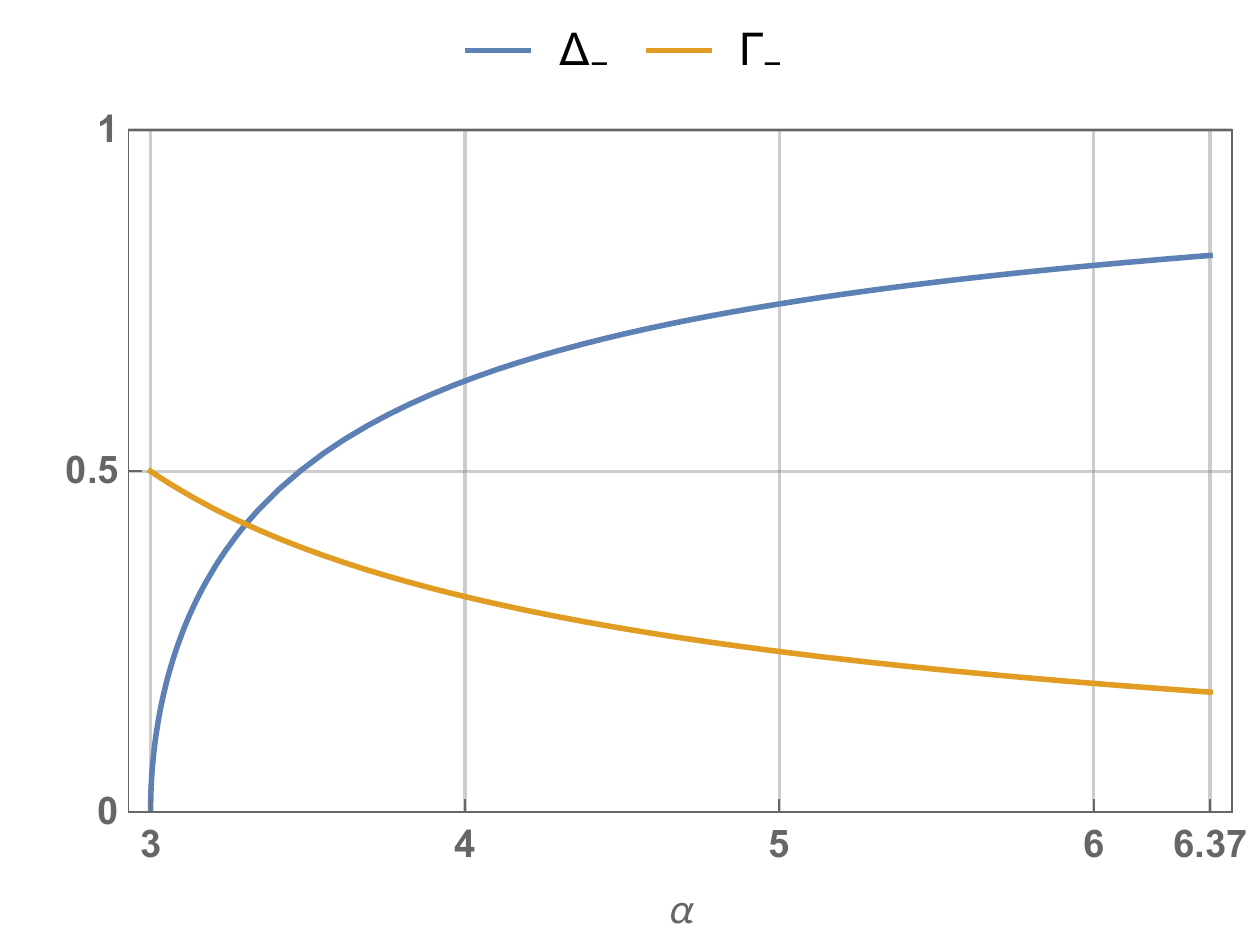}
  \caption{$\Delta_-$ (blue curve) and $\Gamma_-$ (orange curve).}
  \label{FigDGm}
\end{subfigure}
\captionsetup{width=.8\linewidth}
\caption{The mass ratios $\Delta_\pm$ and $\Gamma_\pm$ evaluated at $3<\alpha<\alpha_*\approx 6.37$. In the plot (a) the dashed lines denote the points on the $\alpha$-axis where $\Delta_+=1$ and $\Gamma_+=1$, whose values are (from the left to the right) approximately $3.8$, $5.27$, and $5.37$.}
\label{FigDG}
\end{figure}

In Fig.~\ref{FigDG} we plot the mass ratios for $3<\alpha<\alpha_*$. Fig.~\ref{FigDGp} shows that with $\gamma_+$ corresponding to a positive $\omega_1$ axion is lighter than inflaton if $\alpha<(5+\sqrt{33})/2\approx 5.37$, whereas beyond this point axion becomes heavier. Gravitino (with
 $\gamma_+$) is slightly lighter than inflaton in the range $3.8\lessapprox\alpha\lessapprox 5.27$, whereas  outside this range gravitino becomes heavier.~\footnote{The point $\alpha=(5+\sqrt{33})/2\approx 5.37$ can be found by solving $\Delta_+=1$ that yields a quadratic equation for $\alpha$, while the points $\alpha\approx 3.8$ and $\alpha\approx 5.27$ are found numerically by solving a  quartic equation coming from $\Gamma_+=1$.}

In the case of $\gamma_-$ (see Fig.~\ref{FigDGm}), i.e. a negative $\omega_1$, both axion and gravitino are lighter than inflaton. As we already showed, $x_-$ is the global minimum of the potential even when $\alpha>\alpha_*$, so that the mass ratios $\Delta_-$ and $\Gamma_-$ can be extrapolated for large values of $\alpha$ as
\begin{equation}
    \lim_{\alpha\rightarrow\infty}\Delta_-=1~,~~~\lim_{\alpha\rightarrow\infty}\Gamma_-=0~,
\end{equation}
i.e. the axion mass approaches the inflaton mass, while the gravitino mass slowly vanishes for large $\alpha$.

\section{Inflation}

In order to study inflation, let us restore the gravitational constant $\kappa\equiv\sqrt{8\pi G}=M_P^{-1}$. We choose the K\"ahler potential and the chiral field $T$ to be dimensionless, whereas the superpotential has the mass dimension three, $[W]=M^3$.  It follows that $[\lambda]=[\mu]=M^3$ and $[\omega_1]=M^6$, where $[\ldots]$ stands for the mass dimension of the corresponding quantity. We also set $[g\xi]=M^0$ and $[\phi]=[\varphi]=M$.

It is convenient to express the FI constant $g\xi$ in terms of the cosmological constant $V_0$ by using Eq.~\eqref{V0_alpha>3} and the general $x$-solution \eqref{gammapm}. Restoring $\kappa$ results in the potential
\begin{multline}
    V=V_0+\kappa^2\left(\frac{\gamma}{2}\right)^\alpha\frac{|\mu|^{2(\alpha-1)}}{\omega_1^{\alpha-2}}\left[\frac{\alpha-3}{4}e^{\sqrt{2\alpha}\kappa\varphi}+\frac{\alpha-5}{\gamma}e^{(\alpha-1)\sqrt{\frac{2}{\alpha}}\kappa\varphi}+\frac{\alpha^2-7\alpha+4}{\alpha\gamma^2}e^{(\alpha-2)\sqrt{\frac{2}{\alpha}}\kappa\varphi}\right.-\\
    \left.-\frac{\alpha(\gamma+2)^2}{4\gamma^2}+\frac{(\gamma+2)(3\gamma+14)}{4\gamma^2}-\frac{4}{\alpha\gamma^2}\right]~.\label{V_inf}
\end{multline}
In what follows we neglect the cosmological constant $V_0$.

We use the standard definitions of the slow-roll parameters,
\begin{equation}
    \epsilon\equiv\frac{1}{2\kappa^2}\left(\frac{V'(\varphi)}{V(\varphi)}\right)^2~,~~~\eta\equiv\frac{1}{\kappa^2}\frac{V''(\varphi)}{V(\varphi)}~.
\end{equation}
Inflation ends when $\epsilon=1$  that translates into the value of the inflaton field at the end of inflation, $\varphi_f$. The scalar spectral index and the tensor-to-scalar ratio are related to the slow-roll parameters as
\begin{equation}
    n_s=1+2\eta_i-6\epsilon_i~,~~~r=16\epsilon_i~,\label{nsrdef}
\end{equation}
respectively. Here the subscript $i$ means evaluation at the initial value of the inflaton, $\varphi_i$ 
 i.e., at the horizon crossing. The number of e-foldings between $\varphi_i$ and $\varphi_f$ is given by
\begin{equation}
    N_e=\kappa^2\int^{\varphi_i}_{\varphi_f}d\varphi\frac{V}{V'}~.\label{Nedef}
\end{equation}

Another important observable is the amplitude of scalar perturbations given by
\begin{equation}
    A_s=\frac{\kappa^4V(\varphi_i)}{24\pi^2\epsilon_i}~.\label{As}
\end{equation}

According to the PLANCK data (2018), the observed values of $n_s$, $r$, and $A_s$ are 
 \cite{Akrami:2018odb}
\begin{gather}
    n_s=0.9649\pm 0.0042~{\rm (68\% CL)}~,~~~r<0.064~{\rm (95\%CL)}~,\label{nsr_obs}\\
    \log(10^{10}A_s)=2.975\pm 0.056~{\rm (68\% CL)}~\Rightarrow~A_s\approx 1.96\times 10^{-9}~\label{As_obs}.
\end{gather}

In our models, $n_s$ and $r$ depend only on $\alpha$ and ${\rm sgn}(\omega_1)$ (and not on the value of $\omega_1$) which determine the shape of the scalar potential. The observed value of $A_s$ ($\sim 10^{-9}$) can be used to fix the composite parameter $|\mu|^{2(\alpha-1)}/\omega_1^{\alpha-2}$ that is related to the inflaton mass via Eq.~\eqref{mvarphi^2}.

\begin{figure}
\centering
\begin{subfigure}{.5\textwidth}
  \centering
  \includegraphics[width=1\linewidth]{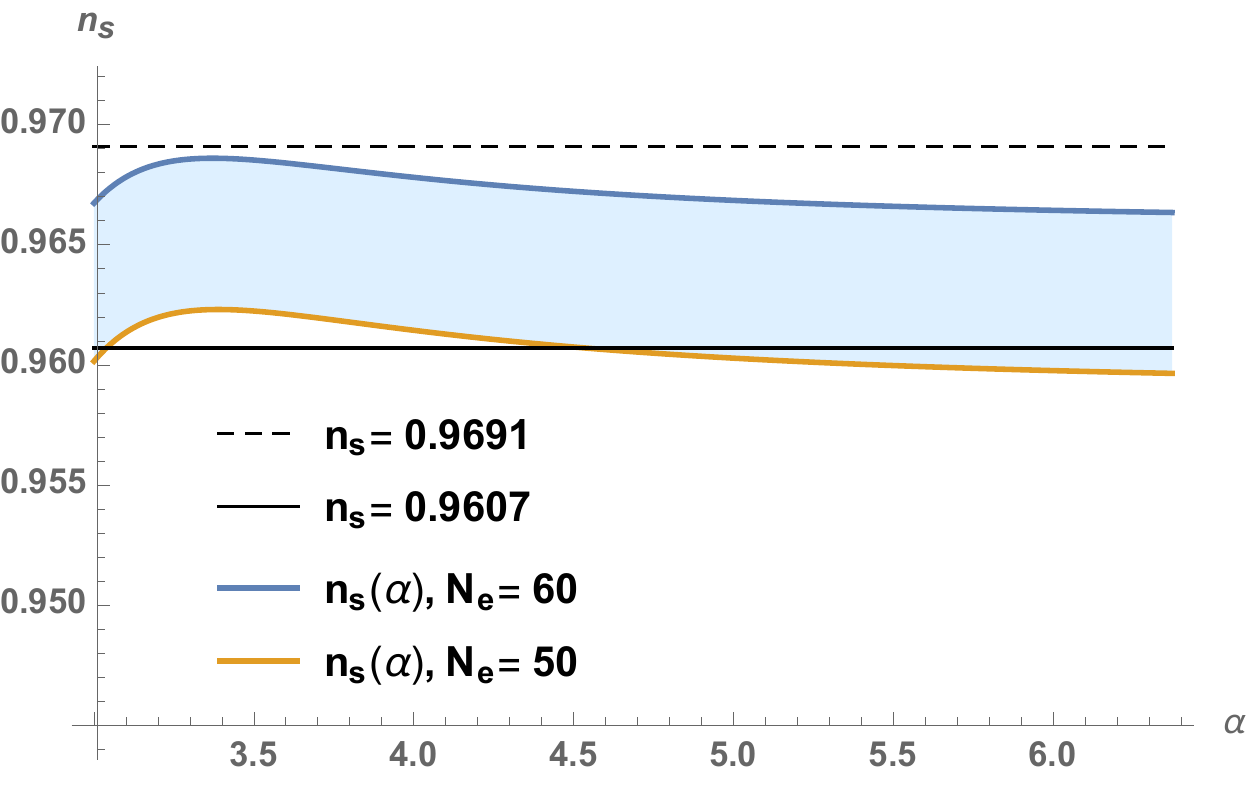}
  \caption{$\omega_1>0$}
  \label{Fignsap}
\end{subfigure}%
\begin{subfigure}{.5\textwidth}
  \centering
  \includegraphics[width=1\linewidth]{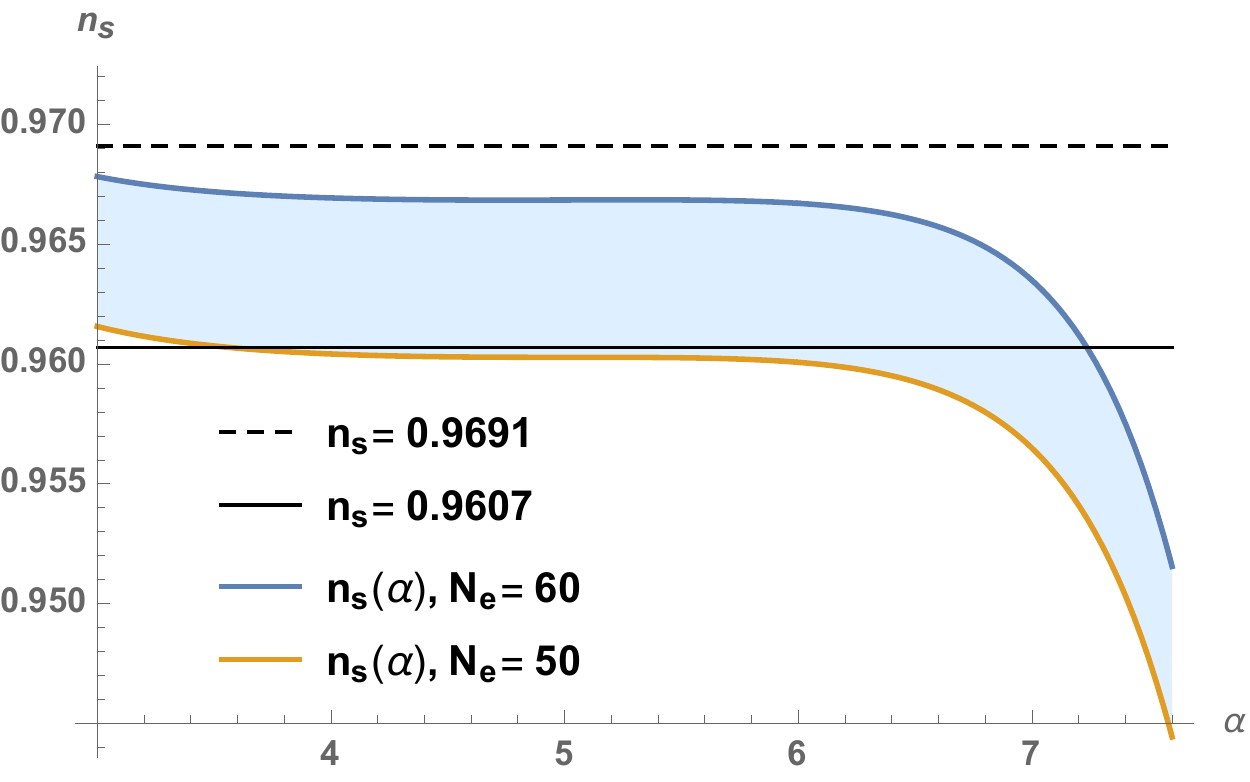}
  \caption{$\omega_1<0$}
  \label{Fignsam}
\end{subfigure}
\captionsetup{width=.8\linewidth}
\caption{The tilt $n_s$ as a function of $\alpha$ for positive and negative $\omega_1$, and $50\leq N_e\leq 60$. The values $n_s=0.9691$ and $n_s=0.9607$ are the upper and lower observational limits ($68\%$CL), respectively.}
\label{Fignsa}
\end{figure}

First, we numerically evaluate $n_s$ as a function of $\alpha$ for $N_e=50$ to $60$. The results of the evaluation are presented in Fig. \ref{Fignsa}. Fig. \ref{Fignsap} shows the tilt $n_s(\alpha)$ evaluated for a positive $\omega_1$ and $3<\alpha<\alpha_*$, while Fig. $\ref{Fignsam}$ shows the tilt $n_s(\alpha)$ evaluated for a negative $\omega_1$ and $3\leq\alpha\leq 7.6$. The $\omega_1>0$ case, in part due to its limited domain of validity ($3<\alpha<\alpha_*$), is fully compatible with the observations of the spectral tilt $n_s$. However, in the $\omega_1<0$ case, if $\alpha$ is greater than  the certain value around $7.2$ (let us call this value $\alpha_{\rm max}$), the predicted value of $n_s$ becomes smaller than the lower observational limit $n_s=0.9607$.~\footnote{In fact, the $n_s$ decreases quite rapidly after $\alpha_{\rm max}$. For example, already for $\alpha=10$ we have $n_s\approx 0.3614$.} A more precise value of $\alpha_{\rm max}$ can be derived by finding 
$\varphi_i$ that solves the condition $n_s(\varphi_i)=0.9607$ and substituting this value in 
Eq.~\eqref{Nedef} to solve $N_e(\alpha)=60$. This results in
\begin{equation}
    \alpha_{\rm max}\approx 7.235~.
\end{equation}
Therefore, when $\omega_1<0$ we exclude the models with $\alpha>\alpha_{\rm max}$.

As we show below, the tensor-to-scalar ratio $r$ decreases with increasing $\alpha$ and is always compatible with the limit $r<0.064$.

\subsection{The case $3\leq\alpha\leq\alpha_*$: Starobinsky-like inflation}

Let us divide our models into two classes for $3\leq\alpha\leq\alpha_*$ and $\alpha_*<\alpha\leq\alpha_{\rm max}$, respectively. The reason is that in the range $3\leq\alpha\leq\alpha_*$ the inflationary potential is truly Starobinsky-like and has a single extremum,
namely, the global minimum  and the infinite plateau asymptotically approaching a constant positive height. In contrast,  if $\alpha>\alpha_*$ the potential has a local maximum, which means that we get the hilltop inflationary models.

\begin{figure}
\centering
\begin{subfigure}{.5\textwidth}
  \centering
  \includegraphics[width=0.91\linewidth]{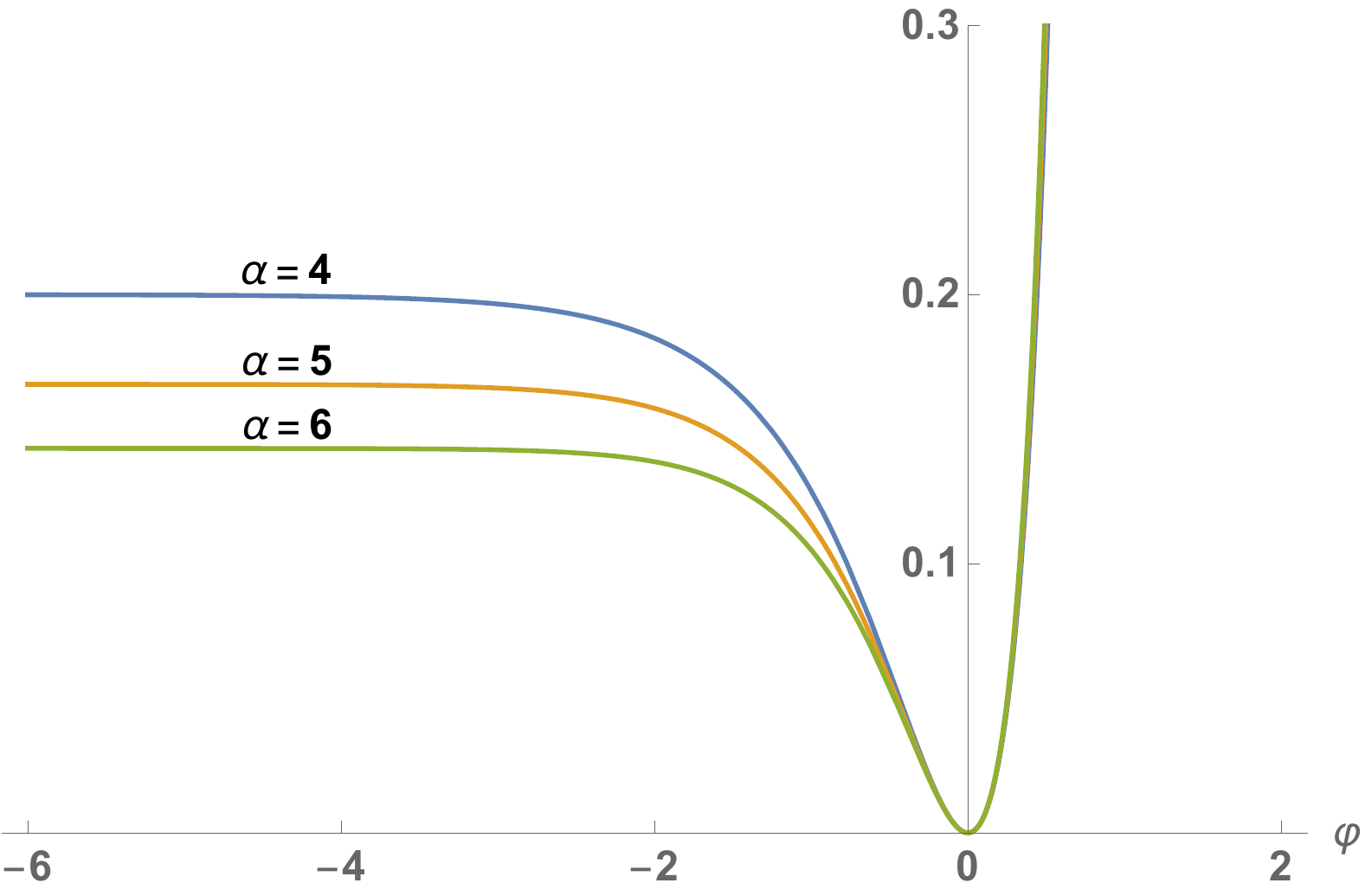}
  \caption{$\omega_1>0$}
  \label{FigV_1a}
\end{subfigure}%
\begin{subfigure}{.5\textwidth}
  \centering
  \includegraphics[width=0.9\linewidth]{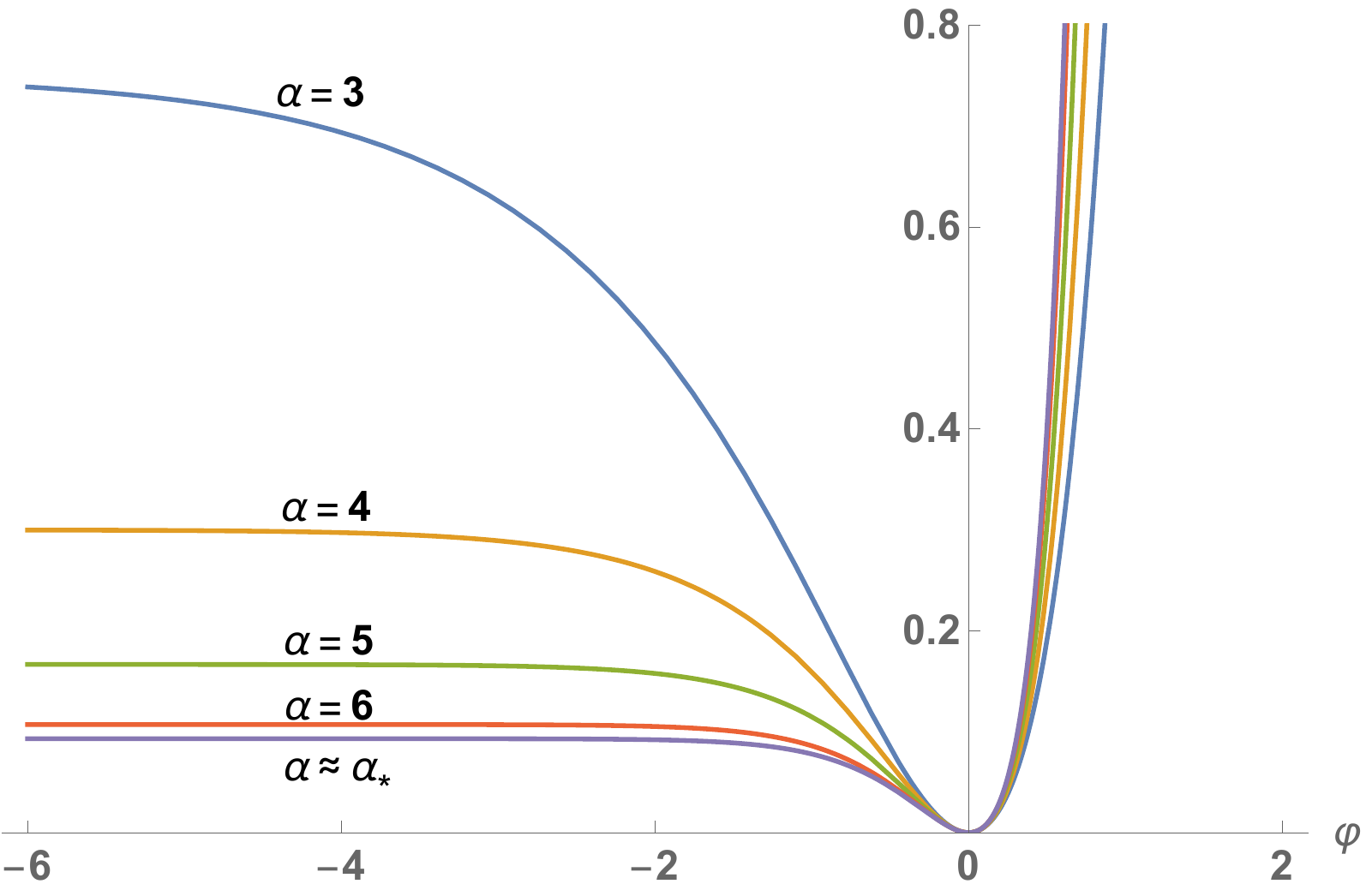}
  \caption{$\omega_1<0$}
  \label{FigV_1b}
\end{subfigure}
\captionsetup{width=.8\linewidth}
\caption{The scalar potentials for both signs of $\omega_1$ and the integer values of $\alpha$ in the range $3\leq\alpha\leq\alpha_*$. The inflaton mass (see Eq.~\eqref{mvarphi^2}) and $\kappa$ are set to be one.}
\label{FigV_1}
\end{figure}

For simplicity, we restrict ourselves to integer $\alpha$, and proceed with calculating the inflationary parameters $n_s$ and $r$ for $3\leq\alpha\leq\alpha_*$ by setting $N_e=55$. In this Subsection, we 
take $\alpha=3,4,5,6$ ($\alpha=3$ is the Starobinsky case) and, in addition, we include the upper limit 
$\alpha=\alpha_*\equiv(7+\sqrt{33})/2$. The results of our numerical calculations of $n_s$ and $r$ are in Table \ref{table_nsr_1}, and the corresponding scalar potentials for the chosen values of $\alpha$ 
are in Fig. \ref{FigV_1}.

\begin{table}[ht]
\centering
\begin{tabular}{|c|c|c|c|c|c|c|c|}
\hline
$\alpha$ & $3$ & \multicolumn{2}{c|}{$4$} & $5$ & \multicolumn{2}{c|}{$6$} & $\alpha_*$\\
\hline
${\rm sgn}(\omega_1)$ & $-$ & $+$ & $-$ & $+/-$ & $+$ & $-$ & $-$ \\
\hline
$n_s$ & $0.9650$ & $0.9649$ & $0.9640$ & $0.9639$ & $0.9634$ & $0.9637$ & $0.9632$\\
\hline
$r$ & $0.0035$ & $0.0010$ & $0.0013$ & $0.0007$ & $0.0005$ & $0.0004$ & $0.0003$\\
\hline
$-\kappa\varphi_i$ & $5.3529$ & $3.5542$ & $3.9899$ & $3.2657$ & $3.0215$ & $2.7427$ & $2.5674$\\
$-\kappa\varphi_f$ & $0.9402$ & $0.7426$ & $0.8067$ & $0.7163$ & $0.6935$ & $0.6488$ & $0.6276$\\
\hline
\end{tabular}
\captionsetup{width=.8\linewidth}
\caption{The predictions for the inflationary parameters ($n_s$, $r$), and the values of $\varphi$ at the horizon crossing ($\varphi_i$) and at the end of inflation ($\varphi_f$), in the case $3\leq\alpha\leq\alpha_*$ with both signs of $\omega_1$. The $\alpha$ parameter is taken to be integer, except of the upper limit $\alpha_*\equiv(7+\sqrt{33})/2$.}
\label{table_nsr_1}
\end{table}

The relation to the amplitude $A_s$ of CMB scalar perturbations in Eq.~\eqref{As} is conveniently described by the composite parameter 
\begin{equation}
    \Lambda^6\equiv\frac{|\mu|^{2(\alpha-1)}}{|\omega_1|^{\alpha-2}}~,
\end{equation}
where $\Lambda$ has units of mass.
When $\omega_1<0$ and $3\leq\alpha\leq\alpha_*$, Eq.~\eqref{As_obs} yields $\Lambda\sim 
10^{101/6}~{\rm GeV}\sim 10^{16.8}~{\rm GeV}$, whereas in the case of $\omega_1>0$ and $3<\alpha<\alpha_*$ we find
\begin{equation}
    \lim_{\alpha\rightarrow 3}\Lambda=0~,~~~\lim_{\alpha\rightarrow\alpha_*}\Lambda=\infty~,
\end{equation}
due to the behavior of $\gamma_+(\alpha)$ (see Eq.~\eqref{gammapm}) in the scalar potential \eqref{V_inf}. Given $\alpha=4,5,6$, the parameter $\Lambda$ is of the order $10^{16.5},10^{16.8},10^{17.5}~{\rm GeV}$, respectively.

The inflaton mass is $m_\varphi\sim 10^{13}~{\rm GeV}$ irrespectively of the choice of $\alpha$ and 
${\rm sgn}(\omega_1)$.

\subsection{The case $\alpha>\alpha_*$: hilltop inflation}

The viable hilltop inflationary models are limited to $\alpha_*<\alpha\leq\alpha_{\rm max}$ with $\alpha_*=(7+\sqrt{33})/2\approx 6.372$ and $\alpha_{\rm max}\approx 7.235$. Let us consider $\alpha=7$, because it is the only integer between $\alpha_*$ and $\alpha_{\rm max}$.

\begin{figure}[t]
\centering
\includegraphics[scale=0.7]{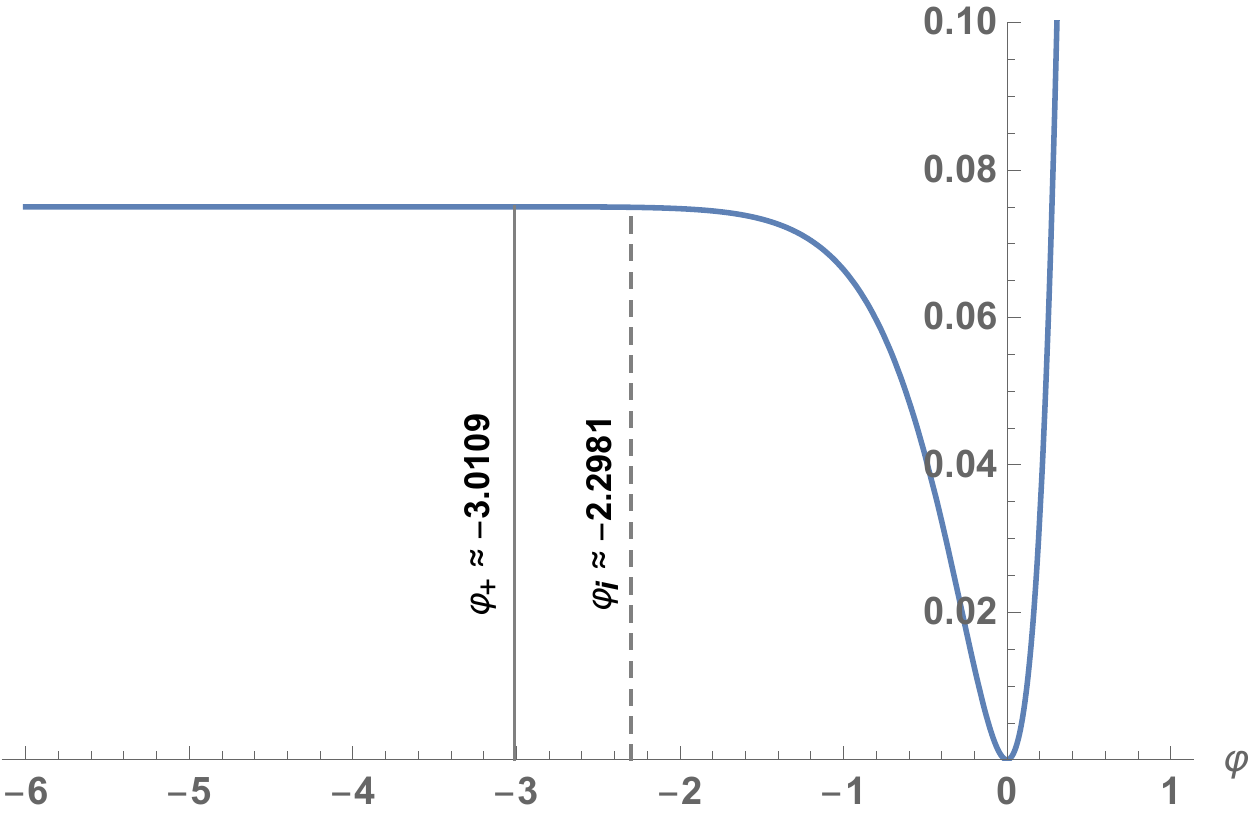}
\captionsetup{width=.8\linewidth}
\caption{The scalar potential \eqref{V_inf} for $\alpha=7$, $\omega_1<0$, and thus $\gamma=\gamma_-$ ($m_\varphi=\kappa=1$). The solid vertical line shows the local maximum 
$\varphi_+$, and the dashed vertical line shows the starting point of inflation $\varphi_i$ when $N_e=60$.}\label{FigV_2}
\end{figure}

Taking $N_e=60$ (for a better fit of $n_s$ with PLANCK data), we calculate the parameters as follows: $n_s\approx 0.9635$, $r\approx 0.0002$, and $\Lambda\sim 10^{16.8}~{\rm GeV}$. The form of the scalar potential is given in Fig.~\ref{FigV_2} where the local maximum $\varphi_+$ and the starting point of inflation $\varphi_i$ are shown.

\subsection{SUSY breaking scale}

Let us parametrize the SUSY breaking scale by the gravitino mass that can be read off from 
Fig.~\ref{FigDG} after taking into account the inflaton mass fixed by the observed value of $A_s$ in 
Eq.~\eqref{As_obs}. For example, if $\omega_1>0$, $m_{3/2}$ ranges from the inflationary scale to arbitrarily high scale (as $\alpha\rightarrow 3$ or $\alpha\rightarrow\alpha_*$). If $\omega_1<0$ and $3\leq\alpha\leq\alpha_{\rm max}$, the gravitino mass is always lower than $m_\varphi$ by at most one order of the magnitude. The exception is the value $\alpha=3$  when $m_{3/2}\geq m_\varphi/2$ as is 
shown in Eq.~\eqref{alpha=3mm}.

In Table \ref{table_SUSY} we provide the explicit values of $m_\varphi$, $m_{t'}$, $m_{3/2}$, $\langle F_T\rangle$, and $\langle D\rangle$ for the integer values of $\alpha$ between $3$ and $\alpha_{\rm max}\approx 7.235$, derived from our models by fixing $A_s$ according to Eq.~\eqref{As_obs}. This fixes $\langle D\rangle=\kappa^{-2}g\xi$ (by using Eqs. \eqref{V0_alpha>3} and \eqref{gammapm}), but it is not enough to fix $\langle F_T\rangle$ in the $\omega_1>0$ case (when $\omega_1<0$, the $\langle F_T\rangle$ identically vanishes except for $\alpha=3$ where it is undetermined). In particular, for $\alpha=4,5,6$, Eq.~\eqref{FTx+} yields
\begin{equation}
    \langle F_T\rangle=\frac{5}{4}\overbar{\mu}\kappa~,~~\langle F_T\rangle=\sqrt{\frac{3}{5}}^{3}\frac{\overbar{\mu}|\mu|}{\sqrt{\omega_1}}\kappa~,~~~\langle F_T\rangle=\frac{7\overbar{\mu}|\mu|^2}{162\omega_1}\kappa~,
\end{equation}
respectively. There is always enough freedom to choose the values of $\langle F_T\rangle$ independently of the parameter $\Lambda^6=|\mu|^{2(\alpha-1)}/|\omega_1|^{\alpha-2}$ that is fixed by the observed amplitude $A_s$.

\begin{table}[ht]
\centering
\hspace*{2.5cm}\begin{tabular}{*{10}{|c}}
\cline{1-9}
$\alpha$ & $3$ & \multicolumn{2}{c|}{$4$} & \multicolumn{2}{c|}{$5$} & \multicolumn{2}{c|}{$6$} & $7$\\
\cline{1-9}
${\rm sgn}(\omega_1)$ & $-$ & $+$ & $-$ & $+$ & $-$ & $+$ & $-$ & $-$\\
\cline{1-9}
$m_\varphi$ & $2.83$ & $2.95$ & $2.73$ & $2.71$ & $2.71$ & $2.53$ & $2.58$ & $1.86$ & \multirow{3}{*}{
    \hspace*{-3mm}$\begin{rcases}
  \\
  \\
  \\
\end{rcases}
\times 10^{13}~{\rm GeV}$
}
\\
\cline{1-9}
$m_{t'}$ & $0$ & $0.93$ & $1.73$ & $2.02$ & $2.02$ & $4.97$ & $2.01$ & $1.56$ & \\
\cline{1-9}
$m_{3/2}$ & $\geq 1.41$ & $2.80$ & $0.86$ & $2.56$ & $0.64$ & $3.91$ & $0.49$ & $0.29$ & \\
\cline{1-9}
$\langle F_T\rangle$ & any & $\neq 0$ & $0$ & $\neq 0$ & $0$ & $\neq 0$ & $0$ & $0$ & \multirow{2}{*}{
\hspace*{-2.6mm}$\begin{rcases}
  \\
  \\
\end{rcases}
\times 10^{31}~{\rm GeV}^2$}\\
\cline{1-9}
$\langle D\rangle$ & $8.31$ & $4.48$ & $5.08$ & $3.76$ & $3.76$ & $3.25$ & $2.87$ & $1.73$ & \\
\cline{1-9}
\end{tabular}
\captionsetup{width=.8\linewidth}
\caption{The masses of inflaton, axion and gravitino, and the VEVs of $F$- and $D$-fields derived from our models by fixing the amplitude $A_s$ according to PLANCK data -- see Eq.~\eqref{As_obs}. 
The value of $\langle F_T\rangle$ for a positive $\omega_1$ is not fixed by $A_s$.}
\label{table_SUSY}
\end{table}

The most important prediction of our models (apart from the existence of the upper limit $\alpha_{\rm max}$) for integer $\alpha$ is the very high SUSY breaking scale parametrized by the superheavy gravitino mass $m_{3/2}$ of the order of $10^{12}$ to $10^{13}$ GeV. For fractional $\alpha$, if $\omega_1>0$,  the SUSY breaking scale can be arbitrarily high as $\alpha$ approaches $3$ or $\alpha_*$.

\section{Conclusion}

In this paper we studied a class of unified models of inflation, spontaneous SUSY breaking, and dark energy (described by the positive cosmological constant) based on the generalized dilaton-axion multiplet coupled to $N=1$ supergravity with the K\"ahler potential and superpotential
\begin{equation}
K=-\alpha\log(T+\overbar{T})~,~~~W=\lambda+\mu T~,\label{Kconcl}
\end{equation}
in the presence of a single vector multiplet with the gauge kinetic function $f=1$.  
 In order to uplift the resulting AdS vacuum, we used the alternative FI term introduced in Refs.~\cite{Cribiori:2017laj,Kuzenko:2018jlz}. This allowed us to get a tunable positive cosmological constant and the D-term contribution to SUSY breaking.

We showed that, unless $\alpha\geq 3$, the scalar potential is unstable. The choice $\alpha=3$ leads to the Starobinsky potential for the dilaton $\varphi$, while the axion direction is flat, i.e. the axion mass has to be generated by quantum corrections. On the other hand, for $\alpha>3$ the axion has a positive non-vanishing mass squared and is automatically stabilized. Once the axion acquires a VEV, those models lead to the effective single-field inflation where inflaton is identified with dilaton. We found that the shape of the potential, and thus the inflationary observables $n_s$ and $r$, are controlled by $\alpha$ and the sign of the real parameter $\omega_1\equiv\overbar{\lambda}\mu+\lambda\overbar{\mu}$, whereas the amplitude of scalar perturbations is related to the value of the composite parameter $\Lambda^6=|\mu|^{2(\alpha-1)}/|\omega_1|^{\alpha-2}$. In particular, when $3\leq\alpha\leq\alpha_*$ ($\alpha_*\approx 6.372$), the derived inflation is of the Starobinsky type where the inflaton rolls down an infinite plateau, while for $\alpha>\alpha_*$ the potential has a local maximum (hilltop).

One of our main results is the upper limit on $\alpha$: by analyzing the dependence of $n_s$ on $\alpha$ (Fig.~\ref{Fignsa}), we found that $\alpha_{\rm max}\approx 7.235$ is the maximum value that can reproduce the observed spectral tilt $n_s=0.9649\pm 0.0042$. More precise observations of $n_s$ may further reduce the value of $\alpha_{\rm max}$.

Another important prediction of our models is the (very) high-scale SUSY breaking, so that for integer $\alpha$ the gravitino mass is roughly of the order of the inflaton mass, $m_{3/2}\sim m_\varphi\sim 10^{13}$ GeV (for fractional $\alpha$, $m_{3/2}$ can be arbitrarily high). In comparison, the scale of the D-term is $\sqrt{|\langle D\rangle|}=\kappa^{-1}\sqrt{g|\xi|}\sim 10^{15.5}$ GeV. We explicitly derived the masses of dilaton, axion and gravitino, together with the SUSY breaking parameters $\langle F_T\rangle$ and $\langle D\rangle$ for $\alpha=3,4,5,6,7$ (see Table \ref{table_SUSY}). It is interesting that the models with a negative $\omega_1$ have the vanishing F-terms $\langle F_T\rangle=0$ (except for $\alpha=3$), so that SUSY is broken purely by the D-term. Those models may be interesting in connection to the universality of scalar masses in the Supersymmetric Standard Model due to the vanishing F-terms, see e.g., Refs.~\cite{Dvali:1997sf,Dudas:2005vv}, though more research is needed in this direction. The axions and gravitinos in our models can be used as the superheavy dark matter along the lines of Refs.~\cite{Addazi:2017ulg,Addazi:2018pbg,Kuzmin:1998kk,Chung:1998ua,Chung:1998zb,Chung:1998rq}.

Although the origin of the alternative FI terms in superstring theory is not clear, the generalized dilaton-axion superfield with the K\"ahler potential given by Eq.~\eqref{Kconcl} with $\alpha=1,2,...,7$ may be derived from M-theory compactified on a $G_2$ manifold \cite{Duff:2010ss,Duff:2010vy,Ferrara:2016fwe}, where the effective $N=1$, $D=4$ supergravity has seven complex scalars parametrizing the $SL(2;\mathbb{R})^7/SO(2)^7$ manifold with the K\"ahler potential
\begin{equation}
K=-\sum_{i=1}^7\log(\Phi_i+\overbar{\Phi}_i)~.
\end{equation}
Then various integer values of $\alpha$ can be obtained by selecting a desired number of $\Phi_i$ superfields and setting the others to be constants. For example, in order  to obtain $\alpha=5$, we can choose $\Phi_1=\Phi_2=...=\Phi_5$ and $\Phi_6=\Phi_7={\rm const}$.

\section*{Acknowledgements}

Y.A. and A.C. were supported by the CUniverse research promotion project of Chulalongkorn University under the grant reference CUAASC. Y.A. was also supported by the Ministry of Education and Science of the Republic of Kazakhstan under the grant reference BR05236322. S.V.K. was supported by the World Premier International Research Center Initiative (WPI), MEXT, Japan, and  the Competitiveness Enhancement Program of Tomsk Polytechnic University in Russia.

\section*{Appendix: $N=1$ supergravity and the alternative FI terms}\label{App1}

The bosonic sector of the standard four-dimensional $N=1$ matter-coupled supergravity reads (in Planck units, $\kappa=1$)~\footnote{A derivation of this action from curved superspace can be found e.g., in Ref. \cite{Wess:1992cp}.}
\begin{equation}
    e^{-1}{\cal L}=\frac{1}{2}R-K_{i\bar{j}}D_m\Phi^i\overbar{D^m\Phi}^j-\frac{1}{4}f^R_{AB}F_{mn}^AF^{B,mn}-\frac{i}{4}f^I_{AB}\tilde{F}_{mn}^AF^{B,mn}-V_F-V_D~,
\end{equation}
whose the F- and D- type scalar potentials are given by
\begin{gather}
    V_F=e^K\left[K^{i\bar{j}}(W_i+K_iW)(\overbar{W}_{\bar{j}}+K_{\bar{j}}\overbar{W})-3|W|^2\right]~,\label{VF}\\
    V_D=\frac{g^2}{2}f_R^{AB}\mathscr{D}_A\mathscr{D}_B~,\label{VD}
\end{gather}
where $K=K(\Phi_i,\overbar{\Phi}_i)$ is the K\"ahler potential depending upon the chiral (complex) scalar fields $\Phi_i$, $F_{mn}^A=\partial_mA_n^A-\partial_nA_m^A+gf^{ABC}A^B_mA^C_n$ is the field strength of the vector fields $A_m^A$, $W=W(\Phi_i)$ is the holomorphic superpotential, $f=f(\Phi_i)$ is the holomorphic gauge kinetic function with $f_R\equiv {\rm Re}f$, $g$ is the gauge coupling, and $\mathscr{D}_A$ are the Killing potentials (the moment maps) of a given gauge group. We use the notation $K^{i\bar{j}}\equiv K_{i\bar{j}}^{-1}$, where $K_{i\bar{j}}\equiv\frac{\partial^2K}{\partial\Phi_i\partial\overbar{\Phi}_j}$, $W_i\equiv\frac{\partial W}{\partial\Phi_i}$, and $f^{AB}\equiv f_{AB}^{-1}$ with $A,B$ as the gauge group indices. The covariant derivatives of the chiral fields are 
\begin{equation}
    D_m\Phi^i=\partial_m\Phi^i-gA_m^AX_A^i~,
\end{equation}
where $X^i_A$ are the Killing vectors of the gauge symmetries.

The potentials \eqref{VF} and \eqref{VD}, as well as the full Lagrangian of $N=1$ supergravity,  are invariant with respect to the K\"ahler-Weyl transformations
\begin{equation}
    K\rightarrow K+\Sigma+\overbar{\Sigma}~,~~~W\rightarrow We^{-\Sigma}~,\label{KWtransform}
\end{equation}
where $\Sigma$ is an arbitrary chiral (super)field.

In Refs.~\cite{Cribiori:2017laj,Kuzenko:2018jlz} the alternative Fayet-Iliopoulos-type terms  in the case of an abelian gauge group were introduced, which do not require gauging the R-symmetry. The FI term with the simplest contribution of the bosonic terms (when using the superspace approach of Ref.~\cite{Wess:1992cp}) reads
\begin{equation}
    {\cal L}_{\rm FI}=-g\xi\int d^2\Theta 2{\cal E}\overbar{\cal P}\left(\frac{{\cal W}^2\overbar{\cal W}^2}{{\cal P}{\cal W}^2\overbar{\cal P}\overbar{\cal W}^2}{\cal DW}\right)+{\rm h.c.}~,\label{superfieldFI}
\end{equation}
where $\xi$ is the real FI constant, $g$ is the $U(1)$ gauge coupling, and $\cal P$ is the chiral projector ${\cal P}\equiv{\cal D}^2-8\overbar{{\cal R}}$, $\overbar{\cal P}\equiv\overbar{\cal D}^2-8{\cal R}$; the (chiral) superfield strength of the vector superfield $\cal V$ is defined by ${\cal W}_\alpha\equiv -\frac{1}{4}\overbar{\cal P}{\cal D}_\alpha {\cal V}$ with ${\cal W}^2\equiv {\cal W}^\alpha {\cal W}_\alpha$. Then the scalar bosonic part of the FI term \eqref{superfieldFI} is just ${\cal L}_{\rm FI}=-eg\xi D$. In order to couple the new FI term to chiral superfields in a K\"ahler-Weyl-invariant manner, one must rescale $\xi\rightarrow e^{-K/3}\xi$ \cite{Aldabergenov:2018nzd,Antoniadis:2018cpq}.

Our models in this paper are described by the potentials
\begin{gather}
\begin{gathered}
K=-\alpha\log(T+\overbar{T})~,\\
W=\lambda+\mu T~,~~~f=1~,\label{KWf}
\end{gathered}
\end{gather}
with complex parameters $\lambda$ and $\mu$. The shift-symmetry $T\rightarrow T+ia$ is not gauged, the Killing potential vanishes, while $V_D$ includes the constant FI contribution (absent in  the standard supergravity) that we use for a dS uplift. After using Eq.~\eqref{KWf} and the parametrization $T=e^{-\sqrt{\frac{2}{\alpha}}\phi}+it$, the full bosonic Lagrangian accounting for the FI contribution takes the form
\begin{equation}
    e^{-1}{\cal L}=\frac{1}{2}R-\frac{1}{2}\partial_m \phi\partial^m\phi-\frac{\alpha}{4}e^{\sqrt{\frac{8}{\alpha}}\phi}\partial_m t\partial^mt-\frac{1}{4}F_{mn}F^{mn}-V_F-V_D~,\label{compL}
\end{equation}
where $V_F$ and $V_D$ are given by Eqs. \eqref{VF0} and \eqref{VD0}, respectively.

\bibliography{bibliography.bib}{}

\providecommand{\href}[2]{#2}\begingroup\raggedright\begin{thebibliography}{10}

\bibitem{Nilles:1983ge}
H.~P. Nilles, ``{Supersymmetry, Supergravity and Particle Physics},''
\href{http://dx.doi.org/10.1016/0370-1573(84)90008-5}{{\em Phys. Rept.}
  {\bfseries 110} (1984) 1--162}.
%%CITATION = PRPLC,110,1;%%.

\bibitem{Aitchison:2007fn}
I.~J.~R. Aitchison, \href{http://dx.doi.org/10.1017/CBO9780511619250}{{\em
  {Supersymmetry in Particle Physics. An Elementary Introduction}}}.
\newblock Cambridge University Press, Cambridge,
2007.
\newblock
%%CITATION = SLAC-R-865;%%.

\bibitem{Ketov:2012yz}
S.~V. Ketov, ``{Supergravity and Early Universe: the Meeting Point of Cosmology
  and High-Energy Physics},''
  \href{http://dx.doi.org/10.1142/S0217751X13300214}{{\em Int. J. Mod. Phys.}
  {\bfseries A28} (2013) 1330021},
\href{http://arxiv.org/abs/1201.2239}{{\ttfamily arXiv:1201.2239 [hep-th]}}.
%%CITATION = ARXIV:1201.2239;%%.

\bibitem{Ketov:2019mfc}
S.~V. Ketov and M.~{\relax Yu}. Khlopov, ``{Cosmological Probes of
  Supersymmetric Field Theory Models at Superhigh Energy Scales},''
\href{http://dx.doi.org/10.3390/sym11040511}{{\em Symmetry} {\bfseries 11}
  no.~4, (2019) 511}.
%%CITATION = 00762,11,511;%%.

\bibitem{Farakos:2013cqa}
F.~Farakos, A.~Kehagias, and A.~Riotto, ``{On the Starobinsky Model of
  Inflation from Supergravity},''
  \href{http://dx.doi.org/10.1016/j.nuclphysb.2013.08.005}{{\em Nucl. Phys.}
  {\bfseries B876} (2013) 187--200},
\href{http://arxiv.org/abs/1307.1137}{{\ttfamily arXiv:1307.1137 [hep-th]}}.
%%CITATION = ARXIV:1307.1137;%%.

\bibitem{Ferrara:2013kca}
S.~Ferrara, R.~Kallosh, A.~Linde, and M.~Porrati, ``{Higher Order Corrections
  in Minimal Supergravity Models of Inflation},''
  \href{http://dx.doi.org/10.1088/1475-7516/2013/11/046}{{\em JCAP} {\bfseries
  1311} (2013) 046},
\href{http://arxiv.org/abs/1309.1085}{{\ttfamily arXiv:1309.1085 [hep-th]}}.
%%CITATION = ARXIV:1309.1085;%%.

\bibitem{Ketov:2014hya}
S.~V. Ketov and T.~Terada, ``{Generic Scalar Potentials for Inflation in
  Supergravity with a Single Chiral Superfield},''
  \href{http://dx.doi.org/10.1007/JHEP12(2014)062}{{\em JHEP} {\bfseries 12}
  (2014) 062},
\href{http://arxiv.org/abs/1408.6524}{{\ttfamily arXiv:1408.6524 [hep-th]}}.
%%CITATION = ARXIV:1408.6524;%%.

\bibitem{Ketov:2014qha}
S.~V. Ketov and T.~Terada, ``{Inflation in supergravity with a single chiral
  superfield},'' \href{http://dx.doi.org/10.1016/j.physletb.2014.07.036}{{\em
  Phys. Lett.} {\bfseries B736} (2014) 272--277},
\href{http://arxiv.org/abs/1406.0252}{{\ttfamily arXiv:1406.0252 [hep-th]}}.
%%CITATION = ARXIV:1406.0252;%%.

\bibitem{Aldabergenov:2016dcu}
Y.~Aldabergenov and S.~V. Ketov, ``{SUSY breaking after inflation in
  supergravity with inflaton in a massive vector supermultiplet},''
  \href{http://dx.doi.org/10.1016/j.physletb.2016.08.016}{{\em Phys. Lett.}
  {\bfseries B761} (2016) 115--118},
\href{http://arxiv.org/abs/1607.05366}{{\ttfamily arXiv:1607.05366 [hep-th]}}.
%%CITATION = ARXIV:1607.05366;%%.

\bibitem{Aldabergenov:2017bjt}
Y.~Aldabergenov and S.~V. Ketov, ``{Higgs mechanism and cosmological constant
  in $N=1$ supergravity with inflaton in a vector multiplet},''
  \href{http://dx.doi.org/10.1140/epjc/s10052-017-4807-8}{{\em Eur. Phys. J.}
  {\bfseries C77} no.~4, (2017) 233},
\href{http://arxiv.org/abs/1701.08240}{{\ttfamily arXiv:1701.08240 [hep-th]}}.
%%CITATION = ARXIV:1701.08240;%%.

\bibitem{Addazi:2017ulg}
A.~Addazi, S.~V. Ketov, and M.~{\relax Yu}. Khlopov, ``{Gravitino and Polonyi
  production in supergravity},''
  \href{http://dx.doi.org/10.1140/epjc/s10052-018-6111-7}{{\em Eur. Phys. J.}
  {\bfseries C78} no.~8, (2018) 642},
\href{http://arxiv.org/abs/1708.05393}{{\ttfamily arXiv:1708.05393 [hep-ph]}}.
%%CITATION = ARXIV:1708.05393;%%.

\bibitem{Aldabergenov:2017hvp}
Y.~Aldabergenov and S.~V. Ketov, ``{Removing instability of inflation in
  Polonyi–Starobinsky supergravity by adding FI term},''
  \href{http://dx.doi.org/10.1142/S0217732318500323}{{\em Mod. Phys. Lett.}
  {\bfseries A91} no.~05, (2018) 1850032},
\href{http://arxiv.org/abs/1711.06789}{{\ttfamily arXiv:1711.06789 [hep-th]}}.
%%CITATION = ARXIV:1711.06789;%%.

\bibitem{Aldabergenov:2018nzd}
Y.~Aldabergenov, S.~V. Ketov, and R.~Knoops, ``{General couplings of a vector
  multiplet in $N=1$ supergravity with new FI terms},''
  \href{http://dx.doi.org/10.1016/j.physletb.2018.07.072}{{\em Phys. Lett.}
  {\bfseries B785} (2018) 284--287},
\href{http://arxiv.org/abs/1806.04290}{{\ttfamily arXiv:1806.04290 [hep-th]}}.
%%CITATION = ARXIV:1806.04290;%%.

\bibitem{Cribiori:2017laj}
N.~Cribiori, F.~Farakos, M.~Tournoy, and A.~van Proeyen, ``{Fayet-Iliopoulos
  terms in supergravity without gauged R-symmetry},''
  \href{http://dx.doi.org/10.1007/JHEP04(2018)032}{{\em JHEP} {\bfseries 04}
  (2018) 032},
\href{http://arxiv.org/abs/1712.08601}{{\ttfamily arXiv:1712.08601 [hep-th]}}.
%%CITATION = ARXIV:1712.08601;%%.

\bibitem{Kuzenko:2018jlz}
S.~M. Kuzenko, ``{Taking a vector supermultiplet apart: Alternative
  Fayet–Iliopoulos-type terms},''
  \href{http://dx.doi.org/10.1016/j.physletb.2018.04.051}{{\em Phys. Lett.}
  {\bfseries B781} (2018) 723--727},
\href{http://arxiv.org/abs/1801.04794}{{\ttfamily arXiv:1801.04794 [hep-th]}}.
%%CITATION = ARXIV:1801.04794;%%.

\bibitem{Abe:2018plc}
H.~Abe, Y.~Aldabergenov, S.~Aoki, and S.~V. Ketov, ``{Massive vector multiplet
  with Dirac-Born-Infeld and new Fayet-Iliopoulos terms in supergravity},''
  \href{http://dx.doi.org/10.1007/JHEP09(2018)094}{{\em JHEP} {\bfseries 09}
  (2018) 094},
\href{http://arxiv.org/abs/1808.00669}{{\ttfamily arXiv:1808.00669 [hep-th]}}.
%%CITATION = ARXIV:1808.00669;%%.

\bibitem{Abe:2018rnu}
H.~Abe, Y.~Aldabergenov, S.~Aoki, and S.~V. Ketov, ``{Polonyi–Starobinsky
  supergravity with inflaton in a massive vector multiplet with DBI and FI
  terms},'' \href{http://dx.doi.org/10.1088/1361-6382/ab0901}{{\em Class.
  Quant. Grav.} {\bfseries 36} no.~7, (2019) 075012},
\href{http://arxiv.org/abs/1812.01297}{{\ttfamily arXiv:1812.01297 [hep-th]}}.
%%CITATION = ARXIV:1812.01297;%%.

\bibitem{Polonyi:1977pj}
J.~Polonyi, ``{Generalization of the Massive Scalar Multiplet Coupling to the
  Supergravity}.'' {Hungary Central Inst. Res. KFKI-77-93 (1977, REC. JUL
  1978), 5 p., unpublished}.

\bibitem{Ellis:1983ei}
J.~R. Ellis, C.~Kounnas, and D.~V. Nanopoulos, ``{Phenomenological SU(1,1)
  Supergravity},''
\href{http://dx.doi.org/10.1016/0550-3213(84)90054-3}{{\em Nucl. Phys.}
  {\bfseries B241} (1984) 406--428}.
%%CITATION = NUPHA,B241,406;%%.

\bibitem{Cremmer:1983bf}
E.~Cremmer, S.~Ferrara, C.~Kounnas, and D.~V. Nanopoulos, ``{Naturally
  Vanishing Cosmological Constant in N=1 Supergravity},''
\href{http://dx.doi.org/10.1016/0370-2693(83)90106-5}{{\em Phys. Lett.}
  {\bfseries 133B} (1983) 61}.
%%CITATION = PHLTA,133B,61;%%.

\bibitem{Ellis:1983sf}
J.~R. Ellis, A.~B. Lahanas, D.~V. Nanopoulos, and K.~Tamvakis, ``{No-Scale
  Supersymmetric Standard Model},''
\href{http://dx.doi.org/10.1016/0370-2693(84)91378-9}{{\em Phys. Lett.}
  {\bfseries 134B} (1984) 429}.
%%CITATION = PHLTA,134B,429;%%.

\bibitem{Ellis:1984bm}
J.~R. Ellis, C.~Kounnas, and D.~V. Nanopoulos, ``{No Scale Supersymmetric
  Guts},''
\href{http://dx.doi.org/10.1016/0550-3213(84)90555-8}{{\em Nucl. Phys.}
  {\bfseries B247} (1984) 373--395}.
%%CITATION = NUPHA,B247,373;%%.

\bibitem{Cecotti:1987sa}
S.~Cecotti, ``{Higher derivative supergravity is equivalent to standard
  supergravity coupled to matter. 1.},''
\href{http://dx.doi.org/10.1016/0370-2693(87)90844-6}{{\em Phys. Lett.}
  {\bfseries B190} (1987) 86--92}.
%%CITATION = PHLTA,B190,86;%%.

\bibitem{Gates:2009hu}
S.~J. Gates, Jr. and S.~V. Ketov, ``{Superstring-inspired supergravity as the
  universal source of inflation and quintessence},''
  \href{http://dx.doi.org/10.1016/j.physletb.2009.03.005}{{\em Phys. Lett.}
  {\bfseries B674} (2009) 59--63},
\href{http://arxiv.org/abs/0901.2467}{{\ttfamily arXiv:0901.2467 [hep-th]}}.
%%CITATION = ARXIV:0901.2467;%%.

\bibitem{Ketov:2009sq}
S.~Ketov, ``{$F(R)$ supergravity},''
  \href{http://dx.doi.org/10.1063/1.3462693}{{\em AIP Conf. Proc.} {\bfseries
  1241} no.~1, (2010) 613--619},
\href{http://arxiv.org/abs/0910.1165}{{\ttfamily arXiv:0910.1165 [hep-th]}}.
%%CITATION = ARXIV:0910.1165;%%.

\bibitem{Starobinsky:1980te}
A.~A. Starobinsky, ``{A New Type of Isotropic Cosmological Models Without
  Singularity},'' \href{http://dx.doi.org/10.1016/0370-2693(80)90670-X}{{\em
  Phys. Lett.} {\bfseries B91} (1980) 99--102}.
[,771(1980)].
%%CITATION = PHLTA,B91,99;%%.

\bibitem{Ellis:2015kqa}
J.~Ellis, M.~A.~G. Garcia, D.~V. Nanopoulos, and K.~A. Olive,
  ``{Phenomenological Aspects of No-Scale Inflation Models},''
  \href{http://dx.doi.org/10.1088/1475-7516/2015/10/003}{{\em JCAP} {\bfseries
  1510} no.~10, (2015) 003},
\href{http://arxiv.org/abs/1503.08867}{{\ttfamily arXiv:1503.08867 [hep-ph]}}.
%%CITATION = ARXIV:1503.08867;%%.

\bibitem{Ellis:2015xna}
J.~Ellis, M.~A.~G. Garcia, D.~V. Nanopoulos, and K.~A. Olive, ``{No-Scale
  Inflation},'' \href{http://dx.doi.org/10.1088/0264-9381/33/9/094001}{{\em
  Class. Quant. Grav.} {\bfseries 33} no.~9, (2016) 094001},
\href{http://arxiv.org/abs/1507.02308}{{\ttfamily arXiv:1507.02308 [hep-ph]}}.
%%CITATION = ARXIV:1507.02308;%%.

\bibitem{Ellis:2017xwz}
J.~Ellis, D.~V. Nanopoulos, and K.~A. Olive, ``{From $R^2$ gravity to no-scale
  supergravity},'' \href{http://dx.doi.org/10.1103/PhysRevD.97.043530}{{\em
  Phys. Rev.} {\bfseries D97} no.~4, (2018) 043530},
\href{http://arxiv.org/abs/1711.11051}{{\ttfamily arXiv:1711.11051 [hep-th]}}.
%%CITATION = ARXIV:1711.11051;%%.

\bibitem{Ellis:2018xdr}
J.~Ellis, B.~Nagaraj, D.~V. Nanopoulos, and K.~A. Olive, ``{De Sitter Vacua in
  No-Scale Supergravity},''
  \href{http://dx.doi.org/10.1007/JHEP11(2018)110}{{\em JHEP} {\bfseries 11}
  (2018) 110},
\href{http://arxiv.org/abs/1809.10114}{{\ttfamily arXiv:1809.10114 [hep-th]}}.
%%CITATION = ARXIV:1809.10114;%%.

\bibitem{Kallosh:2013yoa}
R.~Kallosh, A.~Linde, and D.~Roest, ``{Superconformal Inflationary
  $\alpha$-Attractors},'' \href{http://dx.doi.org/10.1007/JHEP11(2013)198}{{\em
  JHEP} {\bfseries 11} (2013) 198},
\href{http://arxiv.org/abs/1311.0472}{{\ttfamily arXiv:1311.0472 [hep-th]}}.
%%CITATION = ARXIV:1311.0472;%%.

\bibitem{Roest:2015qya}
D.~Roest and M.~Scalisi, ``{Cosmological attractors from $\alpha$-scale
  supergravity},'' \href{http://dx.doi.org/10.1103/PhysRevD.92.043525}{{\em
  Phys. Rev.} {\bfseries D92} (2015) 043525},
\href{http://arxiv.org/abs/1503.07909}{{\ttfamily arXiv:1503.07909 [hep-th]}}.
%%CITATION = ARXIV:1503.07909;%%.

\bibitem{Linde:2015uga}
A.~Linde, ``{Single-field $\alpha$-attractors},''
  \href{http://dx.doi.org/10.1088/1475-7516/2015/05/003}{{\em JCAP} {\bfseries
  1505} (2015) 003},
\href{http://arxiv.org/abs/1504.00663}{{\ttfamily arXiv:1504.00663 [hep-th]}}.
%%CITATION = ARXIV:1504.00663;%%.

\bibitem{Ellis:2018zya}
J.~Ellis, D.~V. Nanopoulos, K.~A. Olive, and S.~Verner, ``{A general
  classification of Starobinsky-like inflationary avatars of SU(2,1)/SU(2)
  $\times$ U(1) no-scale supergravity},''
  \href{http://dx.doi.org/10.1007/JHEP03(2019)099}{{\em JHEP} {\bfseries 03}
  (2019) 099},
\href{http://arxiv.org/abs/1812.02192}{{\ttfamily arXiv:1812.02192 [hep-th]}}.
%%CITATION = ARXIV:1812.02192;%%.

\bibitem{Ellis:2019bmm}
J.~Ellis, D.~V. Nanopoulos, K.~A. Olive, and S.~Verner, ``{Unified No-Scale
  Attractors},''
\href{http://arxiv.org/abs/1906.10176}{{\ttfamily arXiv:1906.10176 [hep-th]}}.
%%CITATION = ARXIV:1906.10176;%%.

\bibitem{Ferrara:2013rsa}
S.~Ferrara, R.~Kallosh, A.~Linde, and M.~Porrati, ``{Minimal Supergravity
  Models of Inflation},''
  \href{http://dx.doi.org/10.1103/PhysRevD.88.085038}{{\em Phys. Rev.}
  {\bfseries D88} no.~8, (2013) 085038},
\href{http://arxiv.org/abs/1307.7696}{{\ttfamily arXiv:1307.7696 [hep-th]}}.
%%CITATION = ARXIV:1307.7696;%%.

\bibitem{Antoniadis:2016aal}
I.~Antoniadis, A.~Chatrabhuti, H.~Isono, and R.~Knoops, ``{Inflation from
  Supergravity with Gauged R-symmetry in de Sitter Vacuum},''
  \href{http://dx.doi.org/10.1140/epjc/s10052-016-4539-1}{{\em Eur. Phys. J.}
  {\bfseries C76} no.~12, (2016) 680},
\href{http://arxiv.org/abs/1608.02121}{{\ttfamily arXiv:1608.02121 [hep-ph]}}.
%%CITATION = ARXIV:1608.02121;%%.

\bibitem{Farakos:2018sgq}
F.~Farakos, A.~Kehagias, and A.~Riotto, ``{Liberated $ \mathcal{N} $ = 1
  supergravity},'' \href{http://dx.doi.org/10.1007/JHEP06(2018)011}{{\em JHEP}
  {\bfseries 06} (2018) 011},
\href{http://arxiv.org/abs/1805.01877}{{\ttfamily arXiv:1805.01877 [hep-th]}}.
%%CITATION = ARXIV:1805.01877;%%.

\bibitem{Antoniadis:2018oeh}
I.~Antoniadis, A.~Chatrabhuti, H.~Isono, and R.~Knoops, ``{The cosmological
  constant in Supergravity},''
  \href{http://dx.doi.org/10.1140/epjc/s10052-018-6175-4}{{\em Eur. Phys. J.}
  {\bfseries C78} no.~9, (2018) 718},
\href{http://arxiv.org/abs/1805.00852}{{\ttfamily arXiv:1805.00852 [hep-th]}}.
%%CITATION = ARXIV:1805.00852;%%.

\bibitem{Cribiori:2018dlc}
N.~Cribiori, F.~Farakos, and M.~Tournoy, ``{Supersymmetric Born-Infeld actions
  and new Fayet-Iliopoulos terms},''
  \href{http://dx.doi.org/10.1007/JHEP03(2019)050}{{\em JHEP} {\bfseries 03}
  (2019) 050},
\href{http://arxiv.org/abs/1811.08424}{{\ttfamily arXiv:1811.08424 [hep-th]}}.
%%CITATION = ARXIV:1811.08424;%%.

\bibitem{Antoniadis:2019hbu}
I.~Antoniadis, J.-P. Derendinger, F.~Farakos, and G.~Tartaglino-Mazzucchelli,
  ``{New Fayet-Iliopoulos terms in ${\mathcal N}=2$ supergravity},''
\href{http://arxiv.org/abs/1905.09125}{{\ttfamily arXiv:1905.09125 [hep-th]}}.
%%CITATION = ARXIV:1905.09125;%%.

\bibitem{Antoniadis:2018cpq}
I.~Antoniadis, A.~Chatrabhuti, H.~Isono, and R.~Knoops, ``{Fayet–Iliopoulos
  terms in supergravity and D-term inflation},''
  \href{http://dx.doi.org/10.1140/epjc/s10052-018-5861-6}{{\em Eur. Phys. J.}
  {\bfseries C78} no.~5, (2018) 366},
\href{http://arxiv.org/abs/1803.03817}{{\ttfamily arXiv:1803.03817 [hep-th]}}.
%%CITATION = ARXIV:1803.03817;%%.

\bibitem{Aldabergenov:2019hvl}
Y.~Aldabergenov, ``{No-scale supergravity with new Fayet-Iliopoulos term},''
  \href{http://dx.doi.org/10.1016/j.physletb.2019.03.068}{{\em Phys. Lett.}
  {\bfseries B795} (2019) 366--370},
\href{http://arxiv.org/abs/1903.11829}{{\ttfamily arXiv:1903.11829 [hep-th]}}.
%%CITATION = ARXIV:1903.11829;%%.

\bibitem{Akrami:2018odb}
{\bfseries Planck} Collaboration, Y.~Akrami {\em et~al.}, ``{Planck 2018
  results. X. Constraints on inflation},''
\href{http://arxiv.org/abs/1807.06211}{{\ttfamily arXiv:1807.06211
  [astro-ph.CO]}}.
%%CITATION = ARXIV:1807.06211;%%.

\bibitem{Dvali:1997sf}
G.~R. Dvali and A.~Pomarol, ``{Supersymmetry breaking with vanishing F terms in
  supergravity theories},''
  \href{http://dx.doi.org/10.1016/S0370-2693(97)00979-9}{{\em Phys. Lett.}
  {\bfseries B410} (1997) 160--166},
\href{http://arxiv.org/abs/hep-ph/9706429}{{\ttfamily arXiv:hep-ph/9706429
  [hep-ph]}}.
%%CITATION = HEP-PH/9706429;%%.

\bibitem{Dudas:2005vv}
E.~Dudas and S.~K. Vempati, ``{Large D-terms, hierarchical soft spectra and
  moduli stabilisation},''
  \href{http://dx.doi.org/10.1016/j.nuclphysb.2005.08.034}{{\em Nucl. Phys.}
  {\bfseries B727} (2005) 139--162},
\href{http://arxiv.org/abs/hep-th/0506172}{{\ttfamily arXiv:hep-th/0506172
  [hep-th]}}.
%%CITATION = HEP-TH/0506172;%%.

\bibitem{Addazi:2018pbg}
A.~Addazi, A.~Marciano, S.~V. Ketov, and M.~{\relax Yu}. Khlopov, ``{Physics of
  superheavy dark matter in supergravity},''
\href{http://dx.doi.org/10.1142/S0218271818410110}{{\em Int. J. Mod. Phys.}
  {\bfseries D27} no.~06, (2018) 1841011}.
%%CITATION = IMPAE,D27,1841011;%%.

\bibitem{Kuzmin:1998kk}
V.~Kuzmin and I.~Tkachev, ``{Matter creation via vacuum fluctuations in the
  early universe and observed ultrahigh-energy cosmic ray events},''
  \href{http://dx.doi.org/10.1103/PhysRevD.59.123006}{{\em Phys. Rev.}
  {\bfseries D59} (1999) 123006},
\href{http://arxiv.org/abs/hep-ph/9809547}{{\ttfamily arXiv:hep-ph/9809547
  [hep-ph]}}.
%%CITATION = HEP-PH/9809547;%%.

\bibitem{Chung:1998ua}
D.~J.~H. Chung, E.~W. Kolb, and A.~Riotto, ``{Nonthermal supermassive dark
  matter},'' \href{http://dx.doi.org/10.1103/PhysRevLett.81.4048}{{\em Phys.
  Rev. Lett.} {\bfseries 81} (1998) 4048--4051},
\href{http://arxiv.org/abs/hep-ph/9805473}{{\ttfamily arXiv:hep-ph/9805473
  [hep-ph]}}.
%%CITATION = HEP-PH/9805473;%%.

\bibitem{Chung:1998zb}
D.~J.~H. Chung, E.~W. Kolb, and A.~Riotto, ``{Superheavy dark matter},''
  \href{http://dx.doi.org/10.1103/PhysRevD.59.023501}{{\em Phys. Rev.}
  {\bfseries D59} (1999) 023501},
\href{http://arxiv.org/abs/hep-ph/9802238}{{\ttfamily arXiv:hep-ph/9802238
  [hep-ph]}}.
%%CITATION = HEP-PH/9802238;%%.

\bibitem{Chung:1998rq}
D.~J.~H. Chung, E.~W. Kolb, and A.~Riotto, ``{Production of massive particles
  during reheating},'' \href{http://dx.doi.org/10.1103/PhysRevD.60.063504}{{\em
  Phys. Rev.} {\bfseries D60} (1999) 063504},
\href{http://arxiv.org/abs/hep-ph/9809453}{{\ttfamily arXiv:hep-ph/9809453
  [hep-ph]}}.
%%CITATION = HEP-PH/9809453;%%.

\bibitem{Duff:2010ss}
M.~J. Duff and S.~Ferrara, ``{Generalized mirror symmetry and trace
  anomalies},'' \href{http://dx.doi.org/10.1088/0264-9381/28/6/065005}{{\em
  Class. Quant. Grav.} {\bfseries 28} (2011) 065005},
\href{http://arxiv.org/abs/1009.4439}{{\ttfamily arXiv:1009.4439 [hep-th]}}.
%%CITATION = ARXIV:1009.4439;%%.

\bibitem{Duff:2010vy}
M.~J. Duff and S.~Ferrara, ``{Four curious supergravities},''
  \href{http://dx.doi.org/10.1103/PhysRevD.83.046007}{{\em Phys. Rev.}
  {\bfseries D83} (2011) 046007},
\href{http://arxiv.org/abs/1010.3173}{{\ttfamily arXiv:1010.3173 [hep-th]}}.
%%CITATION = ARXIV:1010.3173;%%.

\bibitem{Ferrara:2016fwe}
S.~Ferrara and R.~Kallosh, ``{Seven-disk manifold, $\alpha$-attractors, and $B$
  modes},'' \href{http://dx.doi.org/10.1103/PhysRevD.94.126015}{{\em Phys.
  Rev.} {\bfseries D94} no.~12, (2016) 126015},
\href{http://arxiv.org/abs/1610.04163}{{\ttfamily arXiv:1610.04163 [hep-th]}}.
%%CITATION = ARXIV:1610.04163;%%.

\bibitem{Wess:1992cp}
J.~Wess and J.~Bagger, {\em {Supersymmetry and supergravity}}.
\newblock Princeton University Press, Princeton, NJ, USA,
1992.
\newblock
%%CITATION = INSPIRE-350988;%%.

\end{thebibliography}\endgroup
\bibliographystyle{utphys.bst}

\end{document}